\documentclass[apj]{emulateapj}

\usepackage{natbib}
\usepackage{amssymb,amsmath}
\usepackage{graphicx}
\newcommand{\degree}{$^{\circ}$~}
\newcommand{\degreee}{$^{\circ}$}

\newcommand{\kmsec}{km~s$^{-1}$}
\newcommand{\kmsecc}{km~s$^{-1}$ }
\newcommand{\ST}{\emph{STEREO} }
\newcommand{\STA}{\emph{STEREO-A} }
\newcommand{\STB}{\emph{STEREO-B} }
\newcommand{\VEX}{\emph{VEX} }
\newcommand{\MES}{\emph{MESSENGER} }
\newcommand{\WIN}{\emph{Wind} }
\newcommand{\ACE}{\emph{ACE} }

\slugcomment{accepted for publication in \apj}

\shorttitle{Multi-point analysis of multiple ICMEs}
\shortauthors{M\"ostl et al.}

\begin{document}

\title{Multi-point shock and flux rope analysis of multiple interplanetary coronal mass ejections around 2010 August 1 in the inner heliosphere}


\author{C. M\"ostl\altaffilmark{1,2,3}, C.J. Farrugia\altaffilmark{4}, E.~K.~J. Kilpua\altaffilmark{5}, L. K. Jian\altaffilmark{6,7}, Y. Liu\altaffilmark{1,8}, J. Eastwood\altaffilmark{9},  R. Harrison\altaffilmark{10}, D. F. Webb\altaffilmark{11},  M. Temmer\altaffilmark{2}, D. Odstrcil\altaffilmark{7}, J. A. Davies\altaffilmark{10}, T. Rollett\altaffilmark{2,3}, J. G. Luhmann\altaffilmark{1}, N. Nitta\altaffilmark{12}, T. Mulligan\altaffilmark{13}, E. A. Jensen\altaffilmark{14}, R.A. Forsyth\altaffilmark{9}, B. Lavraud\altaffilmark{15,16}, C. A. de Koning\altaffilmark{17}, A. M. Veronig\altaffilmark{2}, A. B. Galvin\altaffilmark{4}, T.~L. Zhang\altaffilmark{3}, and B.~J. Anderson\altaffilmark{18}}

\affil{\altaffilmark{1}Space Science Laboratory, University of California, Berkeley, CA, USA \url{christian.moestl@uni-graz.at}}
\affil{\altaffilmark{2}Kanzelh\"ohe Observatory-IGAM, Institute of Physics, University of Graz, Austria}
\affil{\altaffilmark{3}Space Research Institute, Austrian Academy of Sciences, Graz, Austria}
\affil{\altaffilmark{4}Space Science Center and Department of Physics, University of New Hampshire, Durham, NH, USA }
\affil{\altaffilmark{5}Department of Physics, University of Helsinki, Finland}
\affil{\altaffilmark{6}Department of Astronomy, University of Maryland, College Park, MD, USA}
\affil{\altaffilmark{7}NASA Goddard Space Flight Center, Greenbelt, MD, USA}
\affil{\altaffilmark{8}State Key Laboratory of Space Weather, National Space Science Center, Chinese Academy of Sciences, Beijing, China}
\affil{\altaffilmark{9}The Blackett Laboratory, Imperial College, London, UK}
\affil{\altaffilmark{10}RAL Space, Harwell Oxford, Didcot, UK}
\affil{\altaffilmark{11}Institute for Scientific Research, Boston College, Newton, MA, USA}
\affil{\altaffilmark{12}Solar and Astrophysics Laboratory, Lockheed Martin Advanced Technology Center, Palo Alto, CA, USA}
\affil{\altaffilmark{13}Space Science Applications Laboratory, The Aerospace Corporation, El Segundo, CA, USA}
\affil{\altaffilmark{14}ACS consulting, Houston, TX, USA} 
\affil{\altaffilmark{15}Institut de Recherche en Astrophysique et Plan\'etologie, Universit\'e de Toulouse (UPS), France}
\affil{\altaffilmark{16}Centre National de la Recherche Scientifique, UMR 5277, Toulouse, France}
\affil{\altaffilmark{17}NOAA/SWPC, Boulder Colorado, USA}
\affil{\altaffilmark{18}The Johns Hopkins University Applied
Physics Laboratory, Laurel, MD, USA}




\begin{abstract}


We present multi-point in situ observations of a complex sequence of coronal mass ejections (CMEs) which may serve as a benchmark event for numerical and empirical space weather prediction models. On 2010 August 1, instruments on various space missions \emph{(Solar Dynamics Observatory/ Solar and Heliospheric Observatory/Solar-TErrestrial-RElations-Observatory)} monitored several CMEs originating within tens of degrees from solar disk center. We compare their imprints on four widely separated locations, spanning 120\degree  in heliospheric longitude, with radial distances from the Sun ranging from \emph{MESSENGER} (0.38 AU) to \emph{Venus Express} (\emph{VEX}, at 0.72 AU) to \emph{Wind}, \emph{ACE} and \emph{ARTEMIS} near Earth and \emph{STEREO-B} close to 1 AU. Calculating shock and flux rope parameters at each location points to a non-spherical shape of the shock, and shows the global configuration of the interplanetary coronal mass ejections (ICMEs), which have interacted, but do not seem to have merged. \emph{VEX} and \emph{STEREO-B} observed similar magnetic flux ropes (MFRs), in contrast to structures at \emph{Wind}. The geomagnetic storm was intense, reaching two minima in the \emph{Dst} index ($\approx -100$~nT), caused by the sheath region behind the shock and one of two observed MFRs. \emph{MESSENGER} received a glancing blow of the ICMEs, and the events missed \emph{STEREO-A} entirely. 
 The observations demonstrate how sympathetic solar eruptions may immerse at least 1/3 of the heliosphere in the ecliptic with their distinct plasma and magnetic field signatures. We also emphasize the difficulties in linking the local views derived from single-spacecraft observations to a consistent global picture, pointing to possible alterations from the classical picture of ICMEs.

\end{abstract}





\section{Introduction}

Interplanetary coronal mass ejections (ICMEs) are massive expulsions of plasma and magnetic flux from the solar corona into interplanetary space and are thought to form enormous bent tubes, with radial scale sizes of tenths of an AU, when they are observed in situ by spacecraft in the solar wind. 
With few exceptions \citep[e.g.,][]{bur81, can97, bot98, liu08, moe09b, moe09, rou10, far11, kil11, ruf12}, these measurements have been largely restricted to single--point observations in space. Since their discovery in the beginning of the 1980s \citep{bur81}, this situation has left researchers wondering about their global configuration, because the extreme undersampling results in too many free parameters to unambiguously determine their shape, geometry and magnetic field topology on a large scale.  The same is true for shock waves in the solar wind that are often driven by the magnetic structure inside the ICME \cite[e.g.,][]{ric93}, where it is in general also unclear how to extrapolate, if possible at all, the global shock shape and structure from parameters obtained only at a single point in space. This is important because the structure of the shock determines how particles are accelerated \cite[e.g.,][]{rea99,liu11}.

Furthermore, for testing real--time space weather prediction models \cite[e.g.,][]{sis06}, it would be highly desirable to pinpoint the exact arrival times of an ICME at several locations in the inner heliosphere. Using these known arrival times as boundary conditions, researchers can fine-tune numerical simulations \cite[e.g.,][]{ods04} or empirical methods \cite[e.g.,][]{gop01,kim07, vrs12}, in order to improve the prediction of the ICMEs shock and magnetic ejecta arrival time at Earth. On a deeper level, we ultimately want to understand the physics of the ICME's motion in the solar wind, which was found to be dependent on their heliospheric environment \cite[e.g.,][]{tem11, kil12b}. Single-spacecraft in situ observations can constrain ICME propagation models \citep[e.g.,][]{rol12}, but it is often unclear if the principal direction of the ICME at 1 AU is east or west of the observer and how the shock surface and the flux rope ``tube'' look on a global scale, which is needed for forecasting the arrival time with geometrical models approximating these shapes \cite[e.g.,][]{how09b,lug09a,moe11,moe12,dav12}. With the advent of the \emph{Solar-TErrestrial-RElations-Observatory} \citep[\emph{STEREO}, launched in late 2006, ][]{kai08} and the \emph{Solar Dynamics Observatory} (\emph{SDO}, launched in early 2010), a new era has begun where in situ multi-point observations of ICMEs have become available in connection with high-resolution imaging of the solar corona and multi-point imaging of the entire Sun-Earth line.


Historically, because of a lack of systematic measurements, there have been very few in situ observations of ICMEs at large longitudinal or in-ecliptic separations, and those found were serendipitous \citep[e.g.,][]{bur81,bot98}. The \ST mission, consisting of two spacecraft increasingly leading and lagging the Earth at a rate of 22\degree per year, has made some observations of magnetic flux ropes (MFRs; intervals with smoothly rotating magnetic field) inside ICMEs at up to three points in space (including \emph{Wind} or \ACE near Earth), and it was found that the observations often differed greatly, even for small longitudinal separation of a few degrees \cite[see the extensive summary of these recent and historic events by][]{kil11}. This might be expected for MFRs that have a high inclination of their axis to the ecliptic plane, because one spacecraft passes through the center of the flux rope and the other one crosses along its edge \citep{moe09b,moe09c}. \cite{cro96} showed that it is also possible for a high inclination flux rope to have a strongly elongated cross-section.

For flux ropes with a low inclination to the ecliptic, one can probe the distribution of the magnetic field along the tube, and significant differences of the axial orientations and magnetic fluxes have been found at a spacecraft separation of 20-40\degree \citep{far11}, presumably caused by the interaction with a high-speed stream. \cite{kil11} summarized the available multi-point observations of ICMEs by noting that ``the characteristics of ICMEs and the structure of the solar wind they are embedded in varied significantly from event to event''. 

Thus, the common picture of  flux ropes inside ICMEs as  magnetic loops extending from the Sun into interplanetary space likely needs to be revised to include distortions of the axis and the cross-section. In fact, the magnetic field geometry itself is also uncertain; while it is generally thought to consist of helical field lines, other configurations can lead to similar appearances of the magnetic field in situ \citep{jac09, alh11}, so this matter is also open to further inquiry.

Throughout this paper, we stick to definitions given by \cite{rou11rev}. We define an ``ICME'' as the full interval of solar wind signatures which has changed as compared to the background wind due to the presence of the eruption. This includes the shock, the sheath region of piled-up and compressed plasma behind the shock and the magnetic ejecta, often driving the shock at 1~AU. Subclasses of ejecta are MFRs, which are defined by an interval of smoothly rotating magnetic field in combination with enhanced total field strength. If the interval additionally fulfills low proton temperature they are called ``magnetic clouds'' \citep[MCs; ] []{bur81}. \cite{lep06} also defined ``magnetic cloud like'' (MCL) structures which have the general characteristics of  MCs, but their field rotation is not clear enough to put them into one of the eight categories for MCs defined by \cite{bot98} and \cite{mul98}. \cite{moe10} and \cite{woo11} have found a case where the MCL signature was definitely caused by the spacecraft crossing a leg of an ICME seen in heliospheric images, as assumed by \cite{mar07}.

All the above ideas are made even more complicated by the fact that coronal mass ejections (CMEs) often tend to erupt in rather quick succession, a feature which was originally observed for flares and thus called ``sympathetic'' flaring \citep[see][]{sch11,toe11}. In such a case, one eruption is causally related to the other. In the interplanetary medium, given these favorable initial conditions,  ICMEs might run into one another and interact. Especially around solar maximum, with the next one expected for 2013, this could be a rather common situation (e.g., recent periods of strong successive activity include 2011 February and 2012 March). If as a result of the merging process the identification of  the individual ICMEs is impossible, the structure is called a ``complex ejecta'' \citep{bur02}. However, it may also happen that coalescing events retain their individual structure, but the mutual interaction involving energy and momentum transfer between the ICMEs may alter their field and plasma structure in distinct ways \citep[e.g.,][]{far04, lug09a, liu12}. In general, solar wind streams which show two speed maxima (either CME--CIR or CME--CME) are also called ``compound streams'' and  they are associated frequently with large geomagnetic storms \citep{bur87}. Thus, these streams are of great interest to space weather research in general, simply because the disturbed solar wind interval is both longer and compression may lead to enhanced and more geo-effective southward pointing magnetic fields \citep[e.g.,][]{dal06,zha07,rou10}.


A series of CMEs took place on 2010 August 1, associated with several flares, filament eruptions, and coronal dimmings, and was one of the first periods of strong activity in the rising phase of the current solar cycle 24. In a very fortunate way, many spacecraft equipped with suitable instruments and far separated in heliolongitude were positioned at different heliospheric distances to study this event chain from Sun to Earth with both imaging and in situ observations. To get the most out of this unprecedented dataset, an international team has formed to discuss the events at workshops in Abingdon (UK), Graz (Austria) and Aberystwyth (UK), all held in the first half of 2011. Our paper belongs to a series of studies that form the output of these workshops in an attempt to characterize these events thoroughly, and to lay the foundation for future studies on these events so to obtain more new insights into the physics of solar eruptions. The solar coronal signatures revealed by the \emph{Atmospheric Imaging Assembly} and \emph{Helioseismic and Magnetic Imager} instruments on \emph{SDO}  were studied in detail by \cite{liu10}, \cite{sch11} and \cite{li11b}. The associated CMEs and their connection to the ICMEs observed at Earth using the \emph{Heliospheric Imager} \citep[\emph{HI}; ][]{eyl09} onboard \emph{STEREO-Ahead} has been made by \cite{har12}. \cite{web12} linked the flows seen with the \emph{Solar Mass Ejection Imager} \cite[\emph{SMEI},][]{eyl03} to in situ observations of MFRs. Our paper is meant to close the circle, and to analyze all the available in situ observations in detail, as part of the most complete data set on a sequence of solar eruptions we are aware of, observed from the Sun end--to--end to 1~AU. 

The available in situ observatories were spread around the inner heliosphere at the beginning of 2010 August in an almost laboratory-like configuration, as seen in Figure~\ref{fig:all}(a). \emph{Venus Express (VEX)}, in a polar orbit around Venus at the time of the event, was situated between \emph{STEREO-Behind} and the Earth, and the \emph{MESSENGER} probe was located between \emph{STEREO-Ahead} and the Earth. Thus this spacecraft configuration covered 150\degree in heliospheric longitude. The event also happened before \emph{MESSENGER's} insertion to an orbit around Mercury in 2011 March. The observations and analysis we carry out sets strong constraints on the ICME's propagation directions and orientations as well as their arrival times, which can serve as a benchmark test for space weather prediction efforts using numerical simulations \citep[carried out for these events by][]{ods12,wu11}. They are also of great value in disentangling the signatures of the first multiple ICME fronts observed end--to--end by \emph{STEREO/HI} \citep{har12}, and put constraints on the kinematics of a CME-CME interaction event \citep{tem12, mar12, liu12}. In this sense, we think that this series of events, observed at extremely high resolution on the Sun by SDO, followed to 1~AU with \emph{STEREO/HI} and \emph{SMEI},  and subsequently observed in the solar wind at multiple locations may serve as a standard event for many future investigations on multiple and interacting ICMEs.


\section{Spacecraft positions and overview of observations}\label{sec:overview}

\begin{figure*}[ht]
\epsscale{1.2}
\plotone{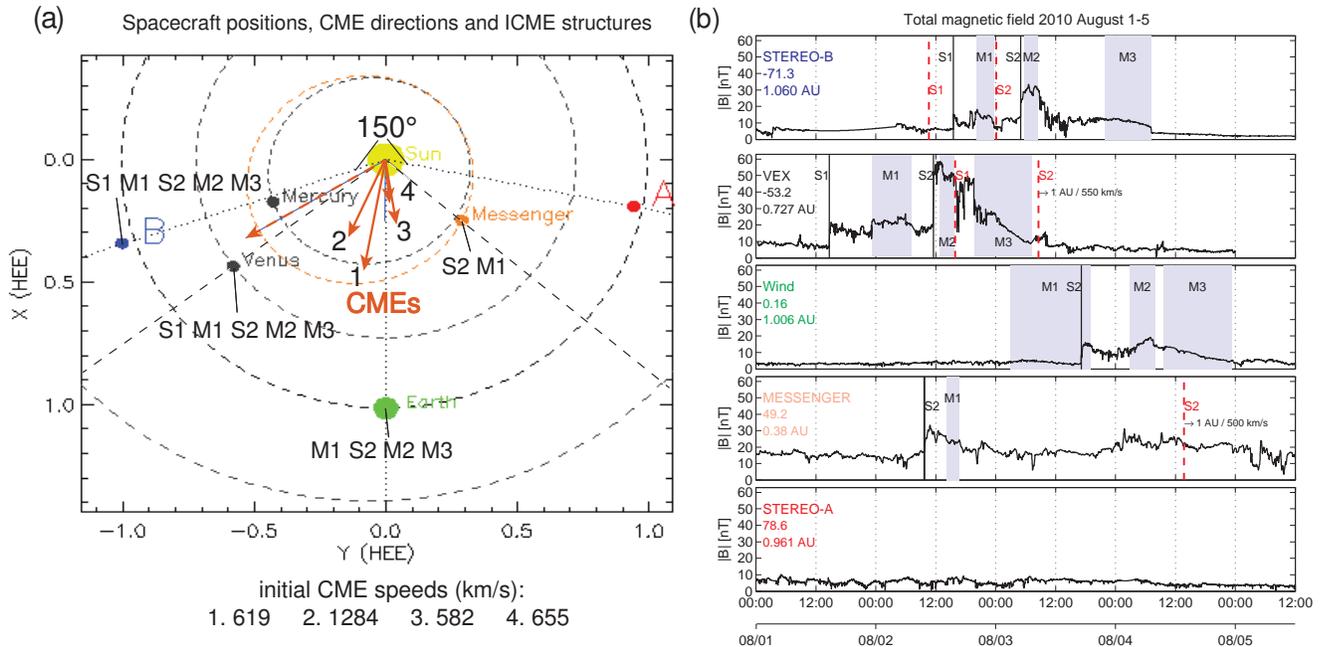}
 \caption{(a): Positions of spacecraft and inner planets on 2010 August 3 00:00 UT in the ecliptic plane (B=\emph{STEREO-B}, A=\emph{STEREO-A}).  The principal directions of one CME on July 30 (dashed) and four CMEs on August 1 (solid, numbered 1--4), are shown by orange arrows, with a shorter arrow indicating a later eruption \citep{har12, ods12}. We also indicated the initial de-projected CME speeds \citep{ods12} for distances $< 15$~$R_{\odot}$. For each spacecraft, the progression of shocks (\emph{S}) and flux ropes (\emph{M}) is indicated. (b): The solar wind total magnetic field observed at five locations in the inner heliosphere ($< 1$~AU) close to the ecliptic plane. From east (\emph{STEREO-B}) to west (\emph{STEREO-A}),  the covered heliospheric longitude is 149.8\degreee, and ICME signatures are observed from \emph{STEREO-B} to \emph{MESSENGER}, corresponding to 120.5\degreee. From top to bottom (or east to west): the total magnetic field at  \emph{STEREO-B}, \emph{Venus Express} (in orbit around Venus), \emph{Wind} (at the L1 point), \emph{MESSENGER}, and \emph{STEREO-A}. Vertical solid lines indicate shock arrival times, and dashed red vertical lines estimated shock arrival times extrapolated to the heliospheric distance of \emph{Wind} ($r_W=1.006$~AU). Shaded intervals indicate well-defined magnetic flux ropes.
}
 \label{fig:all}
\end{figure*}

\begin{table*}[t]\label{tab:positions}
\begin{center}
\caption{Spacecraft Positions}\label{tab:positions}
\begin{tabular}{cccccccc}
\tableline\tableline
Spacecraft & Abbreviation & \emph{r} (AU) & \emph{r} (R$_{\odot}$) & HEE Long.(\degreee) & HEE Lat. (\degreee)  & $B_{max}$ & $B'_{max}$\\
\tableline
\emph{STEREO-Behind} & \emph{B}   &   1.060 &   228.1 &        -71.29       &   -0.29 & 33.2 &  36.2\\
\emph{Venus Express}   &  \emph{V} &  0.727  &  156.2  &          -53.18     &    -0.04  & 58.5 & 34.4\\
\emph{Wind}                 & \emph{W}  &     1.006 &  216.4  &   0.16  &   -0.03 & 19.2 & 19.2\\ 
\emph{MESSENGER}      & \emph{M} &     0.38  &     81.7           &    49.2           &     -5.3 & 33.4 & 6.8 \\
\emph{STEREO-Ahead} & \emph{A} &      0.961 &    206.7           &    78.57 & 0.01  & 10.1 & 9.4\\
\tableline
\end{tabular}\tablecomments{Quoted for 2010 August 3 00:00 UT. The longitude and latitude are given in Heliocentric Earth Ecliptic (HEE) coordinates, with the Earth at 0\degree in both latitude and longitude and the longitude being positive to the west of the Earth. $B_{max}$ is the observed maximum magnetic field strength between 2010 August 1--5, and $B'_{max}$ the scaled field strength to the heliocentric distance of the \WIN spacecraft (see the text).}
\end{center}
\end{table*}


\cite{ods12} identified four significant CMEs on 2010~August~1 with initial ( $< 15~ R_\odot$) directions ranging in longitude from 25\degree east to 8\degree west with respect to the Sun--Earth line. We further quote their results from an elliptical forward modeling technique, which are in general consistent, mostly to within 10\degreee, with other direction-finding techniques \citep[see also][]{tem12,har12}. From Figure~\ref{fig:all}(a), where these directions are indicated as solid red arrows (a shorter arrow indicates a later eruption), it is seen that they  are consistent with the longitudes of the in situ spacecraft, so they can be expected to leave ICME signatures at all of them, aside from \emph{STEREO-A}, if they propagate more or less radially. Another CME on 2010 July 30 (the dashed arrow) was directed mainly towards \emph{STEREO-B}, located at E71. This previous CME is of importance because it influenced the propagation of the August 1 CMEs in the eastern direction \citep{liu12}.

Figure~\ref{fig:all}(a) shows the spacecraft positions on 2010~August~3 00:00 UT, summarized in Table~\ref{tab:positions}, as well as the initial CME directions and speeds, and structures observed in situ at each  spacecraft. The largest separation in heliospheric longitude between \emph{STEREO-Behind}~(``B'' in the figure, at a radial distance from the Sun of $r_B= 1.060$~AU, and most eastward) and \emph{STEREO-Ahead} (``A'', $r_A=0.961$~AU, most westward) was 149.86\degree in Heliocentric Earth Ecliptic (HEE) coordinates. To simplify the analysis we further neglect the small separation angles in HEE latitude between the spacecraft. The \emph{MESSENGER} ($r_M=0.38$~AU) and \VEX ($r_V=0.727$~AU) probes were in extremely favorable positions, each roughly halfway between the respective \emph{STEREO} spacecraft and \emph{Wind}, which was located at the L1 point of the Sun--Earth system ($r_W=1.006$~AU) at the time. Note that \MES in its close orbit to the Sun moves much faster than the other spacecraft. From August 1 to 4 $r_M$ changes from 0.40 to 0.37~AU and the longitude from 42.9\degree to 52.6\degree.

Figure~\ref{fig:all}(b) shows from top to bottom the total magnetic field strength $|\textbf{B}|$ at the locations of \STB \citep{acu08,luh08}, \VEX \citep{zha06}, \WIN \citep{lep95}, \MES \citep{and07}, and \STA \citep{acu08,luh08}, ordered from top to bottom by heliospheric longitude, from east to west. Henceforth we designate by \emph{Mi} (\emph{i}=1, 2, 3)  large--scale magnetic flux ropes and by \emph{Si} (\emph{i}= 1, 2) the shocks (to be discussed in detail in the later sections). 

For the moment we do not imply any associations between the structures observed at each spacecraft. This plot immediately shows that the $|\textbf{B}|$ profile observed by \VEX is qualitatively similar to $|\textbf{B}|$ at \emph{STEREO-B}, at 18$^{\circ}$ separation in longitude, while it looks quite different at \emph{Wind} and \emph{MESSENGER}. At the latter two spacecraft, a single shock is present, in contrast to two clear shocks at \STB and \emph{VEX}. The featureless profile at \STA does not show signs of any shocks or ejecta.

\subsection{Timing Considerations}

It is tempting to think that the shocks \emph{S1} and \emph{S2} in 
Figure~\ref{fig:all}(b) are signatures of two different, extended shock fronts. By this we mean that there actually exist only two large-scale fronts, which the spacecraft sample at different points in space and time.  We can justify this by estimating the arrival times of the shocks if the spacecraft were all positioned at the same heliocentric distance, in our case the heliocentric distance of the \emph{Wind} spacecraft ($r_W$=1.006~AU), which we further use as a reference point. 

To this end, we plot the extrapolated arrival times at $r_W$ of both shocks as red vertical dashed lines in Figure~\ref{fig:all}(b), by using constant speeds of $V_{S1B}=460$~\kmsecc for \emph{S1}, and  $V_{S2B}=600$~\kmsecc for \emph{S2} at \emph{STEREO-B},  which are the shock normal speeds at \STB (derived in Section 3). For \VEX there is no high resolution plasma data available, so we assume this speed is $V_{S2V}=550$ \kmsec, a mean value between the results for \emph{Wind} and \emph{STEREO-B}. While the initial speed of the most eastward CME2 is $\approx 1200$~\kmsec, \cite{tem12} showed that there was a strong deceleration due to interaction with CME1 around 30--50 $R_{\odot}$, and that CME2 attained a constant speed well before the heliocentric distance of Venus. It is also not expected that the ICME speed changes much between 0.72~AU and 1~AU \citep{jia08}. Thus,  assuming this speed to be constant from the Venus distance to about 1~AU and beyond is reasonable. For \MES we use $V_{S2M}=500$~\kmsec, similar to the value observed for \emph{Wind}. Again, for \MES there is no high-time resolution plasma data. If we assume some deceleration up to 1~AU, this choice is consistent with the $\approx$~600 \kmsecc initial speeds for both CMEs heading to the west of Earth \citep[CME3/4,][]{ods12}, and similar propagation speeds for these CMEs derived from \emph{STEREO/HI} observations by \cite{har12}. We also assume that the shock surface moves radially away from the Sun. 

Figure~\ref{fig:all}(b) shows a clear progression from east to west for the time-shifted arrivals (see the red dashed lines): \emph{S2} arrived first at \emph{STEREO-B}, then at \emph{VEX}, followed by \WIN (all three within 12 hr), and last at \emph{MESSENGER}, and it is undetected by \emph{STEREO-A}. The extrapolated arrival times at \emph{STEREO-B} (vertical red dashed lines) in Figure~\ref{fig:all} are earlier than the observed ones because \emph{STEREO-B} is at a larger heliocentric distance than \emph{Wind}. The close timing compels us to believe that we are observing the same shock surface at the different locations, and the late \MES observations and the lower shock normal speed at \WIN are expected because for a convex surface, such as an expanding circular front attached to the Sun, the side parts (or ``flank'') will move slower than the front part \citep[also called ``apex'' or ``nose'' of the ICME; cf.][]{moe11, moe12}. We thus deduce qualitatively that the shock surface for \emph{S2} has a convex shape, with a smaller radial distance of the shock front correlated with positive longitude values for a given point in time. This  means that the shock front is in this case not spherical with respect to the Sun. 

There are two effects which we think contribute to this shape. First,  the previous ICME on July 30, which is associated with shock \emph{S1} \citep{liu12}, created a region of depleted density in the solar wind, so the eastern part of the ICMEs originating on August 1 experienced less drag, which led to less deceleration. Secondly, one has to keep in mind that the initial speed of CME2, with an initial direction between E25 and E41, was $\approx 1200$~\kmsecc and thus about double than for the other CMEs (all $\approx 600$~\kmsec).

 The shock \emph{S1} is constrained to the east of Earth and is observed first at \emph{STEREO-B}, then at \emph{VEX}, but neither at \emph{Wind} nor \emph{MESSENGER} so we cannot derive an approximate global shape like for \emph{S2}.

\subsection{Shape of the Shock Fronts}
\begin{figure*}[h]
\epsscale{1.3}
\plotone{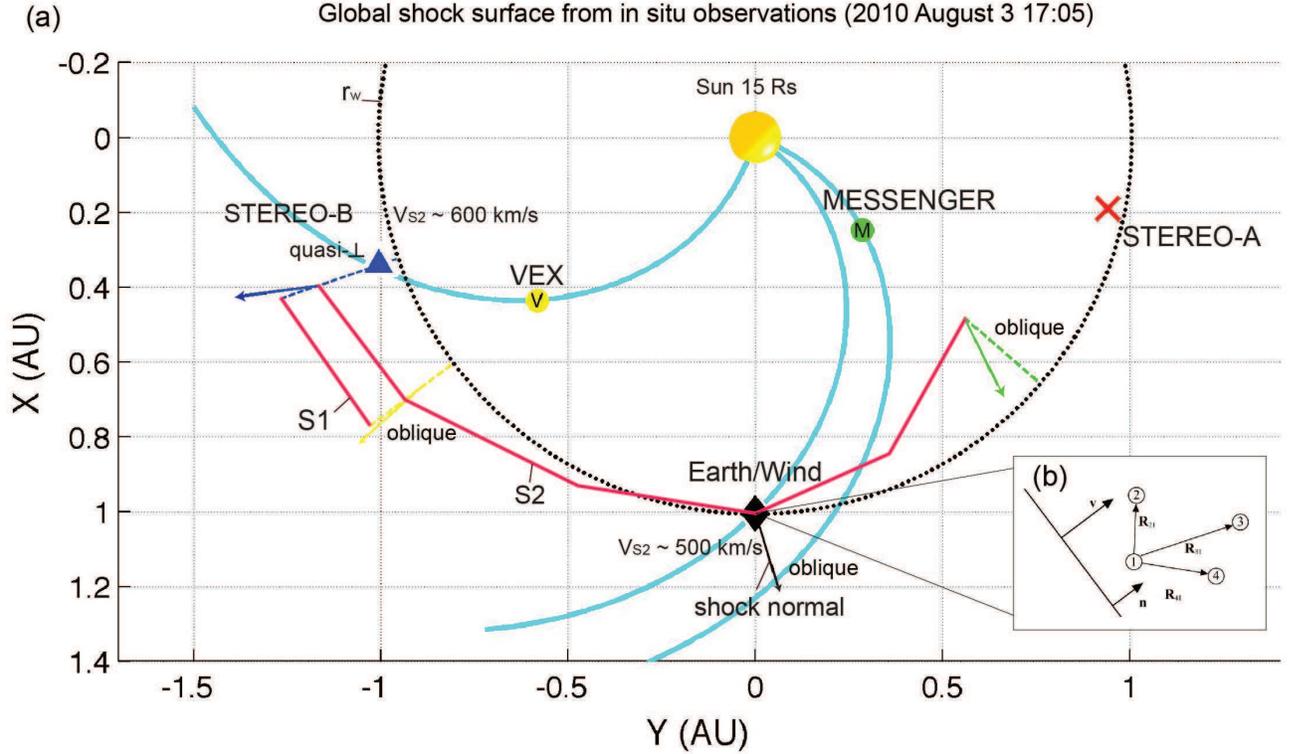}
\caption{(a) Global configuration of two shock fronts \emph{S1} and \emph{S2}, estimated at the time of the shock observation at \WIN (2010 August 3 17:05~UT). The approximate shape of the shocks (red solid lines) is shown by estimating the heliocentric distance of the shocks at each location for the given point in time.  For \emph{S2}, shock normals are drawn as arrows in the color consistent with the spacecraft position, the configuration of the shock is indicated as oblique or quasi-perpendicular. The nominal Parker spiral field is drawn connecting the Sun to each spacecraft  for the upstream conditions of solar wind speed 450~\kmsecc at \STB and \VEX and 400~\kmsecc at \WIN and \emph{MESSENGER}. Note that  \VEX and \STB are approximately connected by the nominal Parker spiral field. (b) The insert shows an illustration of the four-spacecraft method to determine the shock normal orientation near Earth, using \emph{Wind}, \emph{ACE} and \emph{ARTEMIS P1/P2}. A planar shock is measured at four points in space for which the distances to one reference spacecraft have to be known. }\label{fig:global_shock}
\end{figure*}

We have also deduced the shape of the shock front in a more quantitative way. In many situations, this shape is unknown or accessible only through numerical simulations. However, this shape is of great interest for predicting the arrival time of ICMEs as a function of  heliospheric longitude \citep[e.g.,][]{moe12}, and we are especially interested in its appearance for a complex situation involving interacting ICMEs.

In Figure~\ref{fig:global_shock}(a) we plotted the estimated heliocentric distance of the shock (e.g., $r_{S2B}$ for the shock \emph{S2} at \emph{STEREO-B}) along radials from the Sun to each spacecraft for the time the shock is observed at \emph{Wind} ($t_{S2W}$ = 2010 August 3 17:05 UT) using 
\begin{equation}\label{shockestimate}
 r_{S2B}=r_B+V_{S2B} (t_{S2W}-t_{S2B}),
\end{equation}
with $V_{S2B}$ and $t_{S2B}$ being the \emph{S2} shock normal speed and its  arrival time, respectively at \emph{STEREO-B}. Again, we assume radial propagation of the shock. The result is $r_{S2B}=1.2349$~AU, which is considerably larger than $r_w=1.006$~AU. Both  distances would be similar if the shock had the shape of a sphere centered on the Sun.  Similarly, estimating these distances for \emph{VEX} and \emph{MESSENGER} by using similar speeds as for the time-shifts (550 \kmsecc for \VEX and 500 \kmsecc for \emph{MESSENGER}),  we get $r_{S2V}=1.1188$~AU and $r_{S2M}=0.7412$~AU. We may thus plot the shock front \emph{S2} as the red solid line connecting the dots \citep[cf.][]{bur81} at the snapshot of the shock arrival time at \WIN in Figure~\ref{fig:global_shock}(a), showing clearly the non-spherical shape. Between \VEX and \WIN as well as \WIN and \MES we plotted another point, estimated by eye for a smoother form of the shock surface. 

A similar analysis  was done for shock \emph{S1}, which seems to have a shape closer to spherical than \emph{S2}, but with only two data points it is difficult to discuss the global structure of \emph{S1} so we will further concentrate on \emph{S2}.

As we will discuss in a later section, we note that if \emph{S2} is indeed one large shock surface spanning $> 120$\degree in heliolongitude, it likely is driven by different structures at different locations along the front.

\subsection{Scaling the Magnetic Field Strength}

Additionally, to fully characterize the observed complex flows in a simple fashion, we can deduce how the maximum observed magnetic field strength scales with heliospheric longitude, and again, we need to roughly compensate for the different heliospheric distances  $r$ of the spacecraft. As a proxy, we use an empirical scaling relation for the inner heliosphere by \cite{lei07}, who found that the maximum magnetic field strength $B_0$ inside MCs in the inner heliosphere ($< 1$~AU) depends on radial distance as
\begin{equation}\label{leitnerlaw}
   B_{0}(r) =18.1  r^{-1.64}, 
\end{equation}
with $B_0$ in (nT) and $r$ in (AU).  For example, the peak field strength was observed by \VEX at the front of the flux rope \emph{M2} as $B_{maxV}=58.5$~nT at a radial distance of $r_V$. To estimate the maximum $B'_{maxV}$ scaled to the radial distance of $r_W$, we simply eliminate the factor $18.1$ to obtain
\begin{equation}\label{mylaw}
 B'_{maxV}=B_{maxV}  (r_W/r_{V})^{-1.64},
\end{equation}
yielding $B'_{maxV}= 34.4$~nT. At \emph{STEREO-B} and \emph{MESSENGER}, the peak field (see Table~\ref{tab:positions}) is also detected shortly after \emph{S2}, while it is more delayed to \emph{S2} at \emph{Wind}, our reference location. Taken together, the maximum field strengths estimated at $r_W$ compare as: 36.2~nT (\emph{STEREO-B}), 34.4~nT (\emph{VEX}), 19.2~nT (\emph{Wind}, not scaled), and 6.8~nT (\emph{MESSENGER}). For comparison, the result is 9.4~nT for \emph{STEREO-A}, which did not observe the structure. Thus, the high peak field at \MES is scaled down to normal solar wind levels at 1~AU.  This also means that the shock, if it had an initial component towards \emph{STEREO-A}, is likely to have decayed when it reached 1~AU at its western edge.

Because one gets easily lost in these huge amounts of data, it is useful to keep the basic picture which emerged from this section in mind as a background for the following, more detailed  analysis. We now move on to calculate parameters of the shocks and flux ropes to obtain an even more detailed global picture using the in situ observations.  In addition to parameters discussed so far, these may be useful for more detailed comparisons to numerical simulations.

\section{Shock analysis}

\begin{table*}[h]
\begin{center}
\caption{ Shock Parameters at Each Spacecraft.}\label{tab:shocks}
\begin{tabular}{lcccccccccc}
\tableline\tableline
Spacecraft & Shock   & Long.(\degreee)  & Arrival (UT) &  Normal $\mathbf{n_s}$ (xyz) &  $\theta_s$(\degreee) & $\phi_s$(\degreee) & $V_s$(\kmsec) &  $q$(\degreee) & $M_A$ & $B_d/B_u$  \\
(1) & (2) & (3) & (4) & (5) & (6) & (7) & (8) & (9) & (10) & (11)\\
\tableline
\STB &S1  & -71.3 & Aug  2 15:30 & (-0.998, 0.058, -0.004)& 0 & 177 & 460 & 69 & 1.9 & 2.3\\
\VEX & S1& -53.2 & Aug  1 14:41 & (-0.761, 0.263, -0.594)&  -36 & 160     & \ldots  & 56 & \ldots & 2.3      \\
\tableline
\STB &S2  & -71.3 &Aug  3 05:00 & (-0.981, 0.194, 0.025) & 1 & 169 & 600  & 79 & 3.1 & 1.7 \\
\VEX &S2 & -53.2 &Aug  2 11:30 & (-0.981, -0.094, 0.182)     &  10 & 185   & \ldots  & 54 & \ldots & 2.6 \\
\WIN &S2  & 0.2  &Aug 3 17:05 &(-0.831, -0.259, 0.492) &  29 & 197 & 497 & 56 & 4.0 & 3\\
\emph{Wind, ACE, ARTEMIS P1/P2} &S2 & $\sim$ 0 &Aug 3 17:05 &(-0.907, -0.218, 0.361) &  21 & 194  & 511 &  42 & 5.0 & 2.6\\
\MES &S2 & 49.2 & Aug 2 09:43 &  (-0.794, 0.345, -0.501) & -30 & 157 & \ldots & 44 & \ldots  & 1.4 \\
\tableline
\end{tabular}
\end{center}
\tablecomments{Column 1: observing spacecraft. Column 2: name of the shock. Column 3: longitude of the spacecraft in HEE coordinates. Column 4: shock arrival time. Column 5: shock normal in GSE coordinates for near-Earth spacecraft, and approximated GSE coordinates for the other spacecraft (i.e., RTN rotated by 180\degree in the RT plane for \MES and \STB and VSO coordinates for \emph{VEX}). Column 6: latitude of the shock normal angle with respect to the ecliptic plane. Column 7:  longitude of the  shock normal measured from GSE $+X$ ($0^{\circ}$) to GSE $+Y$ ($90^{\circ}$). Column 8: shock normal speed. Column 9: angle between the shock normal and the upstream field (also known as $\theta_{Bn}$). Column 10: Alfv\'{e}n Mach number. Column 11: ratio between the downstream and upstream field strengths.}
\end{table*}

\begin{figure*}[h]
\epsscale{0.8}
\plotone{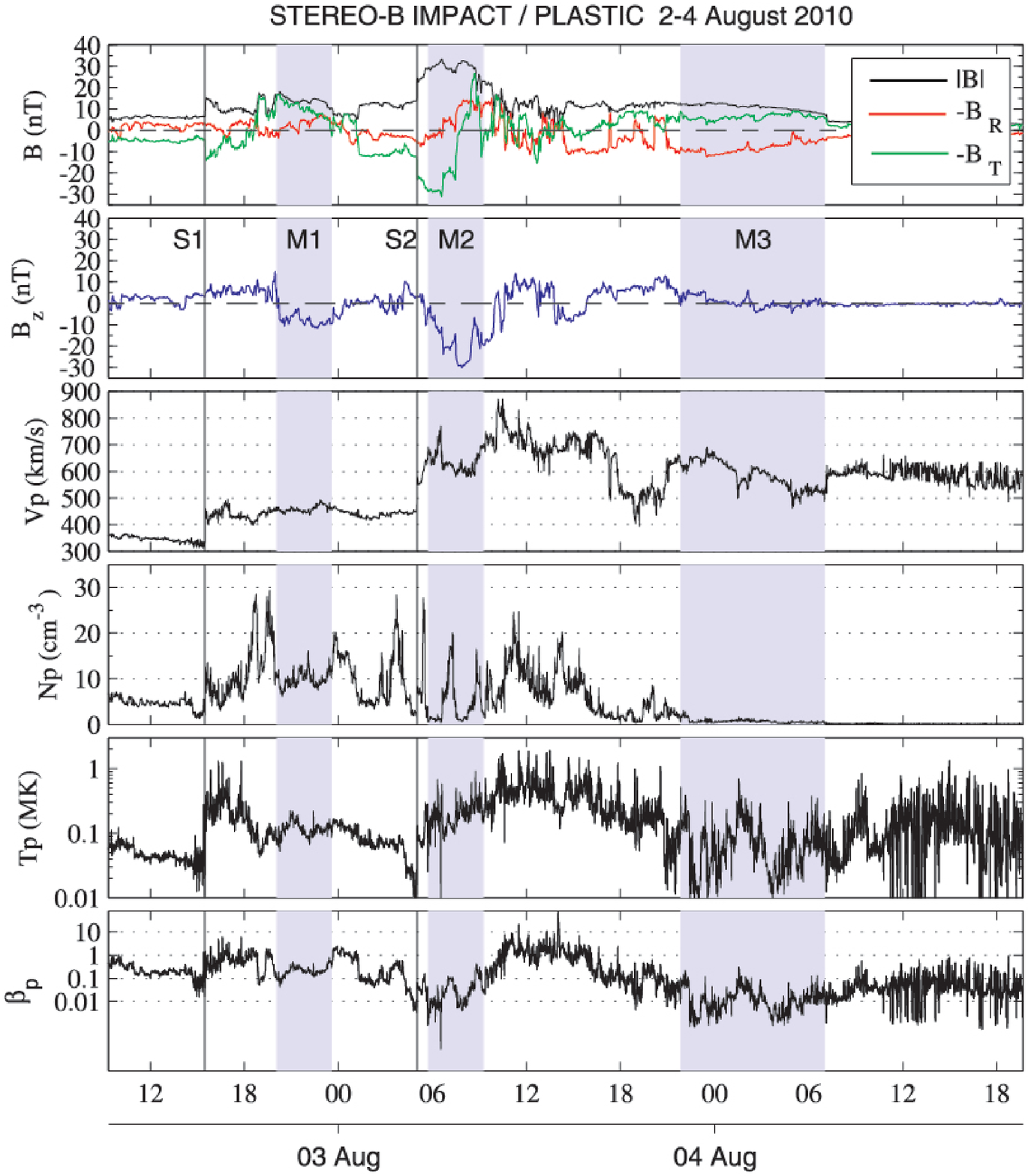}
 \caption{From top to bottom: magnetic field components $-B_R$ (red), $-B_T$ (green), and total field magnitude (black); magnetic field $B_z$ component;  proton bulk speed; proton number density; proton temperature; and proton $\beta$. Two shocks are indicated by vertical solid lines along \emph{S1/2} and magnetic flux ropes are indicated by shaded regions \emph{M1/2/3}.  } \label{fig:stereob}
\end{figure*}

The aim of this section is to analyze the observed shocks in detail and to relate these findings to the global shape and structure of the shocks, giving us a rare opportunity to do this on a global scale with multiple spacecraft \citep[e.g.,][]{bur81}. 
Parameters of interplanetary shocks, such as the shock normal orientation, the configuration of the shock (perpendicular, oblique or parallel), and the Mach number, can be obtained by using the co-planarity theorem, which follows from the general Rankine--Hugoniot jump conditions of magnetohydrodynamics \citep[e.g.,][]{kiv95, bur95, sch98}. The theorem states that the upstream (index $u$, or ``unshocked'' or ``undisturbed'' region of the solar wind) and downstream (index $d$, or ``disturbed'') vectors of the interplanetary magnetic field (IMF) and the shock normal $\mathbf{n}_s$ all lie in the same plane, and the normal can be obtained with

\begin{equation}
\label{equ:coplanar}
      \mathbf{n}_s=\frac{(   \mathbf{B}_u - \mathbf{B}_d)  \times (\mathbf{B}_u \times \mathbf{B}_d)}{\left|    (\mathbf{B}_u - \mathbf{B}_d)  \times (\mathbf{B}_u \times \mathbf{B}_d) \right|}. 
\end{equation}

For near-Earth spacecraft, we will express the resulting shock normal in geocentric solar ecliptic coordinates with $X$ pointing toward the Sun, $Z$ toward ecliptic north and $Y$ completing the right handed triad (thus pointing towards solar east). Also, we will quote angles for the shock normal $\theta_s$ (latitude) and $\phi_s$ (longitude) which are defined as such that $\theta_s$ is measured with respect to the ecliptic plane ($\theta_s=0$\degreee, and ecliptic north = +90\degreee) and $\phi_s$ is $0$\degree along GSE $X$ and $90$\degree along GSE $Y$. For \emph{STEREO-A/B} and \emph{MESSENGER}, we will simply rotate the result, first given in radial-tangential-normal (RTN) coordinates, which are defined as $R$ pointing away from the Sun and $T$ towards solar west, by 180\degree in the solar equatorial plane, meaning we replace $R$ with $-R$ and $T$ with $-T$, so the orientation of the coordinate system is similar to GSE. For \emph{VEX}, Venus Solar Orbital (VSO) coordinates are used, which are defined similar to GSE but for the orbit of Venus. We are thus neglecting small differences in orbital inclination between the Earth, Venus, and the solar equatorial plane (see also Table~\ref{tab:positions}). Typical errors for the shock normal with the co-planarity method are in the order of 10\degreee--20\degreee, so the intrinsic errors are definitely larger than the systematic error we impose by our simplified treatment of the coordinate systems. 

If solar wind bulk velocity vectors are available (only for the \emph{Wind} spacecraft), one can use the conservation of mass flux across the shock surface to obtain the shock normal speed $V_s$ with
\begin{equation}
\label{equ:shockspeed}
     V_s= \frac{N_d \mathbf{V}_d -N_u \mathbf{V}_u}{N_d-N_u} \cdot \mathbf{n_s}  
\end{equation}
where $N_d$, $N_u$, $\mathbf{V_d}$, and $\mathbf{V_u}$ are the downstream and upstream solar wind densities and velocities, respectively \citep{kiv95}. In the case when no plasma bulk velocity vectors are available but its norm is known (for \emph{STEREO-A/B}), we can still estimate the shock speed  if we suppose that upstream and downstream speeds are radials \citep{bur95} and replace $\mathbf{V_u}$ and $\mathbf{V_d}$ by their respective norms $V_u$ and $V_d$ in Equation~(\ref{equ:shockspeed}). For \emph{VEX} and \emph{MESSENGER} we will derive the shock normal only because no high resolution plasma data are available.

The Alfv\'{e}nic Mach number defines the strength of the shock and it is calculated as $M_A = |V_{sr} - V_u|/V_A$, where $V_A$ is the Alfv\'{e}n speed in the upstream region and $V_{sr}$ the shock radial speed ($V_{sr}=V_s/n_{sx}$). The angle $q$ (often called $\theta_{Bn}$ in the literature, but we would like to avoid this because in this paper $\theta$ denotes latitudes for flux rope and shock orientations), gives the angle between the upstream field $\mathbf{B}_u$ and $\mathbf{n}_s$ and defines the shock as quasi-parallel when $q < 45$\degree and quasi-perpendicular if $q > 45$\degreee. Shocks with $q$-values around 45\degree are called oblique. The parameter $q$ defines the mechanisms of particle acceleration at shocks \citep{rea99}, and shocks propagating inside ICMEs have higher values of $q$ \citep{ric10perp}. At 1~AU, for a radially propagating spherical shock, $q$ is expected to be close to $45$\degree because it is the angle of the Parker spiral field to the radial direction.

\subsection{Application to \emph{STEREO-B}, \emph{VEX} and \emph{MESSENGER}}

We will now discuss the application to the  \emph{STEREO-B}  shocks and summarize briefly the results at  \emph{VEX} and \emph{MESSENGER}. Figure \ref{fig:stereob} shows the plasma and magnetic field measurements by \emph{PLASTIC} \citep{gal08} and \emph{IMPACT} \citep{luh08} aboard \emph{STEREO-B}.

\textbf{Shock \emph{S1}:} We used an interval August 2 05:08--05:53 UT. The results for this method are $\textbf{n}_{s1b}=(-0.998,   0.058, -0.004)$ (in coordinates where $R$ is replaced by $-R$ and $T$ by $-T$), or $\theta_{s1B}=0^\circ$ / $\phi_{s1B}=177^\circ$, and $q_{1B}=69^\circ$. The Alfv\'{e}nic Mach number was 1.9 and the ratio of the upstream to downstream field was 2.3, both values pointing to a moderate to weak shock.

\textbf{Shock \emph{S2}:}
For interval August 3 04:40--05:20 UT we get $\textbf{n}_{s2B}=(0.981,  -0.194,  0.025)$, or $\theta_{s2B}=1^\circ$ / $\phi_{s2B}=169^\circ$, and $q_{2B}=79^\circ$. Here, $M_A$ was 3.1 and the $B_d/B_u$=1.7, again in a moderate regime of shock strength.  For a compilation of these results, see Table~\ref{tab:shocks}.

\begin{figure*}[T]
\epsscale{1}
\plotone{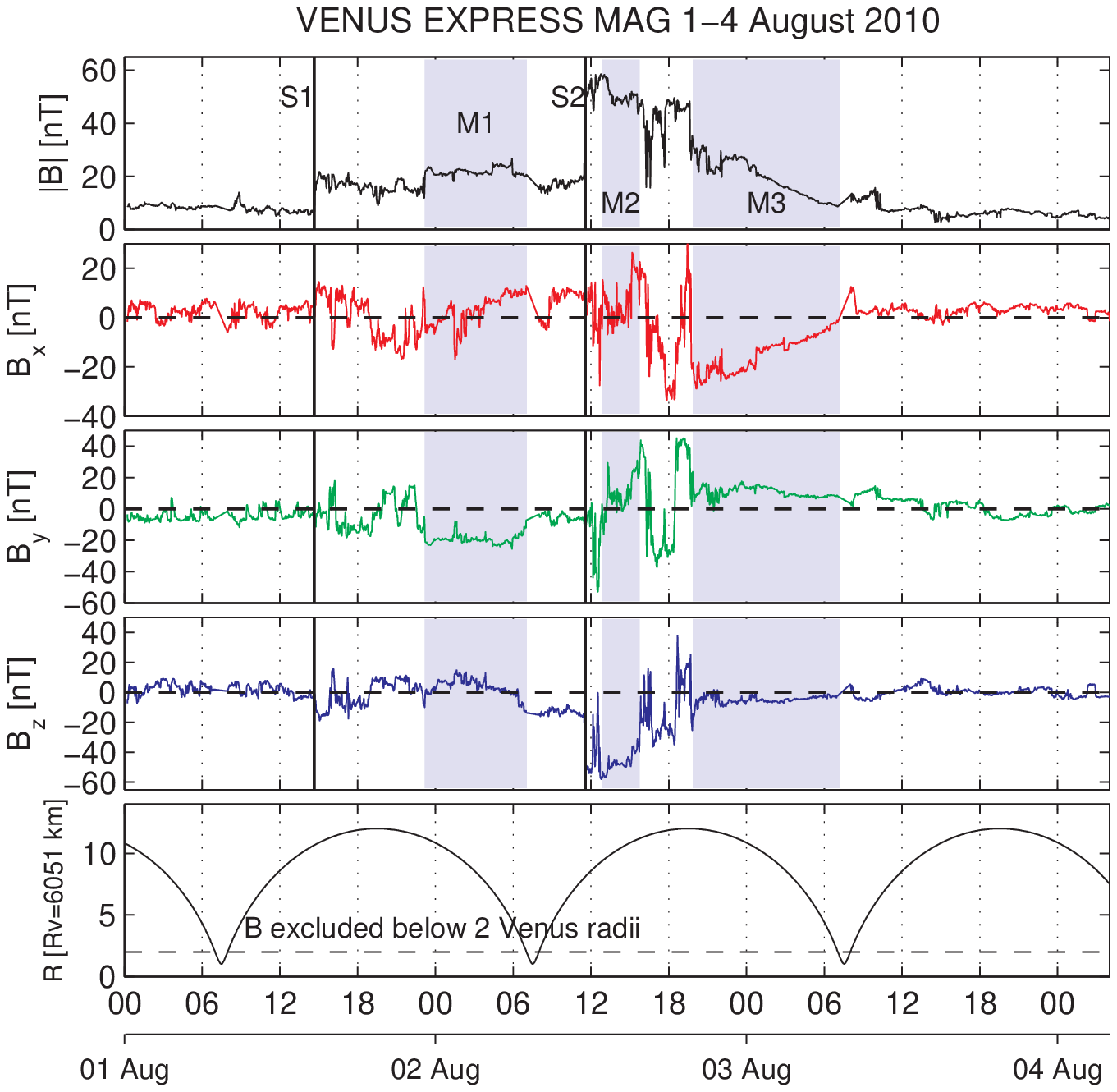}
 \caption{From top to bottom: total magnetic field and field components observed by \emph{VEX/MAG}, in Venus Solar Orbital (VSO) coordinates (similar to GSE, but for the orbit of Venus). The bottom panel shows the radial distance of the spacecraft to the center of Venus. We linearly interpolated the magnetic field over times \VEX approached Venus closer to two Venus radii to see only the solar wind magnetic field. }\label{fig:vex}
\end{figure*}

Figure~\ref{fig:vex} shows the magnetic field measured by \emph{VEX/MAG} \citep{zha06}. We linearly interpolated over intervals when \emph{VEX} was closer than two Venus radii to the planet to avoid plotting field strengths strongly exceeding those in the solar wind when the spacecraft flies through Venus' induced magnetosphere.  The application of the co-planarity theorem shows that both shock normals were close to radial from the Sun. The shocks were of oblique type,  and of  moderate to weak shock strength.

\begin{figure*}[T]
\epsscale{1}
\plotone{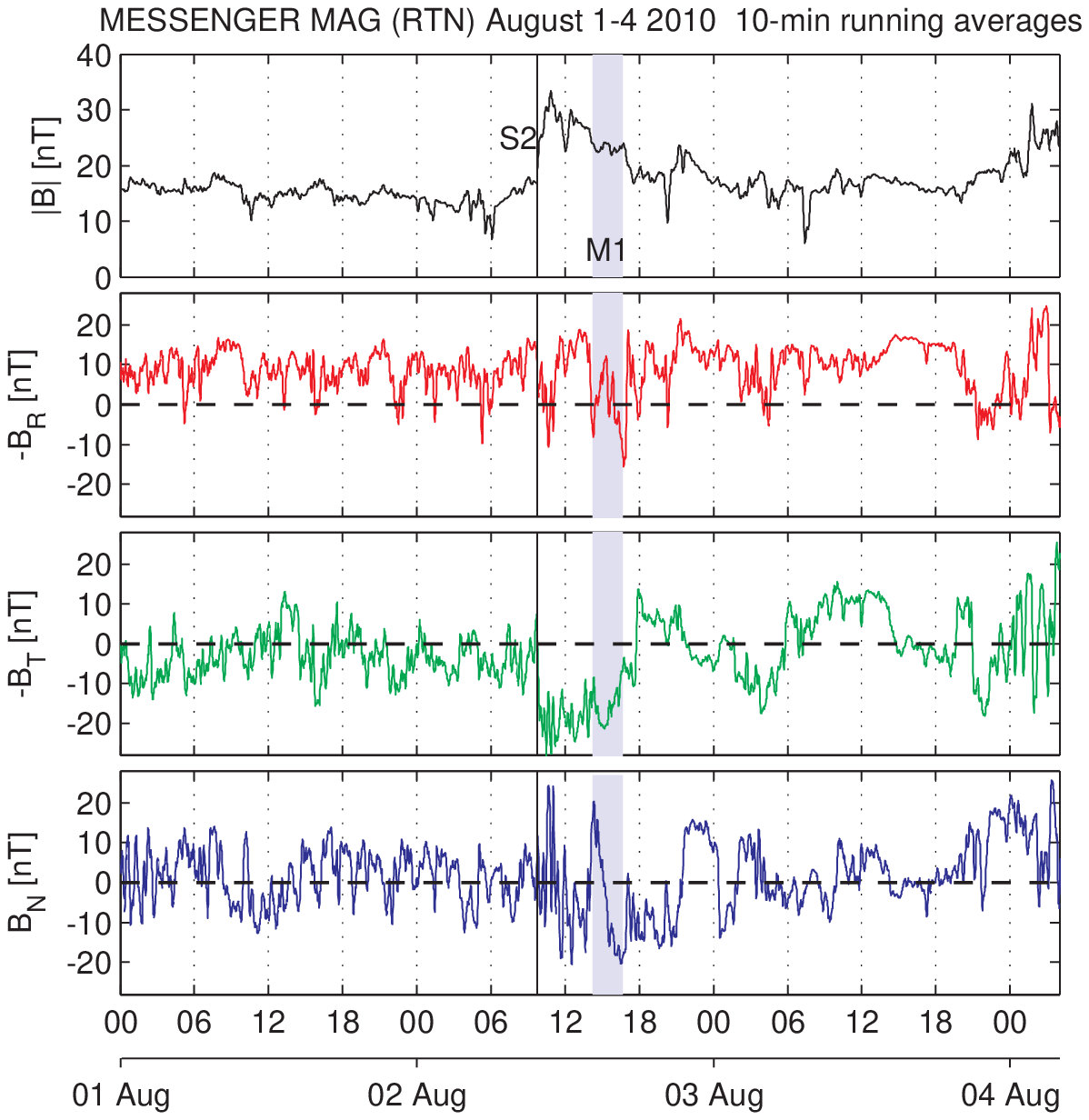}
 \caption{Total magnetic field and magnetic field components observed by \emph{MESSENGER}. For easier comparison to the components observed by the other spacecraft, component $R$ is replaced by $-R$ (pointing radially to the Sun) and component $T$ by $-T$ (pointing towards solar east). The solid vertical line marks the shock \emph{S2}, and the shaded interval a magnetic flux rope (\emph{M1}).}
 \label{fig:messenger}
\end{figure*}

 \emph{MESSENGER} observed a  single  shock on August 2 9:43 UT as seen in Figure~\ref{fig:messenger}. The derived parameters indicate a very weak shock strength with $B_d/B_u=1.4$ and the strongest deviation of the shock normal from the radial direction. The low strength at this western location could also indicate that if the shock \emph{S2} had an initial component directed towards \emph{STEREO-A}, it could have already decayed when it reached this spacecraft, as noted previously when we estimated how the field strength scales with heliospheric distance.

\subsection{Near-Earth Space}

\begin{figure*}[H]
\epsscale{1}
\plotone{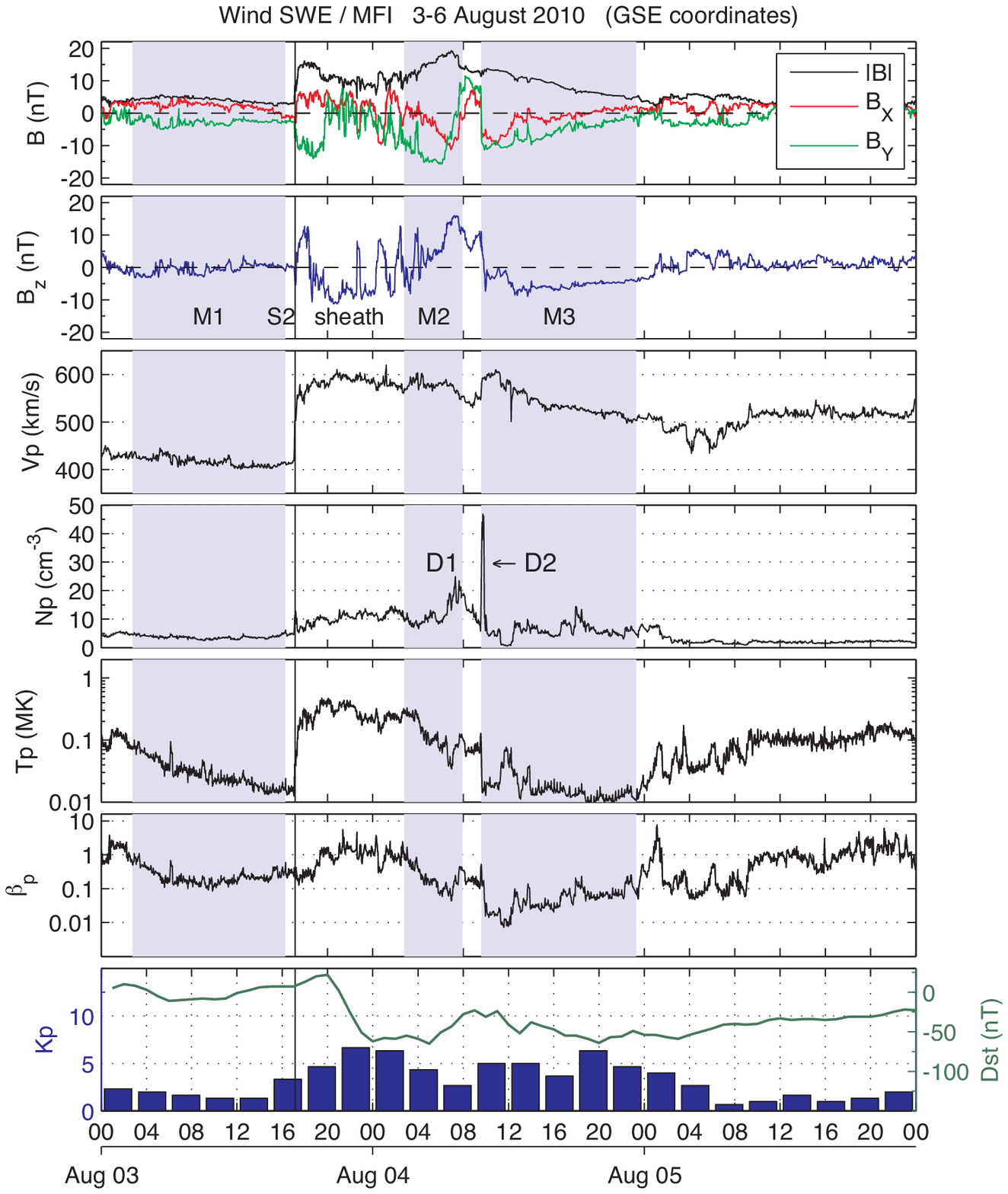}
 \caption{Plasma and magnetic field observations by the \emph{Wind} spacecraft. From top to bottom: magnetic field strength (black) and GSE components $B_X$ (red) and $B_Y$ (green), magnetic field $B_z$ (blue), proton bulk speed, proton density, proton temperature, proton $\beta$, and the $Dst$ (green line) and $Kp$ (blue bars) indices for geomagnetic activity. The solid vertical line denotes the shock \emph{S2}, and shaded intervals mark magnetic flux ropes. Two proton density enhancements are labeled as \emph{D1/2}.  }\label{fig:wind}
\end{figure*}
 
\WIN is currently in orbit around the Lagrangian point L1 and at the time of the event it was located at $R_w=(205, -64, -14)$ Earth radii ($R_E$) in GSE coordinates. Magnetic field \citep[MFI instrument;][]{lep95} and plasma \citep[SWE;][]{ogi95} observations between 2010 August 3 and 4 by the \WIN spacecraft are shown in Figure~\ref{fig:wind}. Because we are interested mainly in the relation of the magnetic field vector to the same observations by the other widely separated spacecraft \VEX and \emph{STEREO-B}, we plot here the magnetic field components in GSE coordinates. Later, when discussing geoeffects, we will use the appropriate GSM coordinates.

At 2010 August 3 17:05 UT, \WIN detected a forward shock that is indicated in Figure~\ref{fig:wind} with a solid line and labelled \emph{S2}. At the shock the solar wind speed increased abruptly from 425~km~s$^{-1}$ to about 505~km~s$^{-1}$.
  Simultaneously, the ion density increased from 5 to 12 cm$^{-3}$, and the magnetic field magnitude from $\sim$3~nT to $\sim$9~nT. Note that this shock shows, when zoomed in, a two-step profile, so the solar wind speed actually increases to $\approx 600$~\kmsecc later in the sheath region. Here, in addition to the co-planarity theorem we were able to use a four--spacecraft method which is based on the different shock arrival times and the locations of the spacecraft \citep{rus83,sch98}. The shock \emph{S2} arrived at \emph{ACE} on August 3 16:55 UT, and at the two \emph{ARTEMIS} \citep[former \emph{THEMIS}]{sib11} spacecraft \emph{P1} on 17:20 UT and \emph{P2} at 17:40 UT, while both were in transit from Earth to lunar orbit.

\textbf{Method (1):}

We used 3~s magnetic field measurements from \emph{Wind} in the interval 16:50--17:20 UT. The  co-planarity approach yielded a shock normal orientation $\textbf{n}=(-0.831, -0.259, 0.492)$ in GSE coordinates (latitude  $\theta_s=28^\circ$; longitude $\phi_s=197^\circ$). 
 The shock speed was $V_s=497$~km~s$^{-1}$, and the shock was oblique with $q=56^\circ$. From $V_A=38$~\kmsec~ being the Alfv\'{e}n speed in the upstream region and $V_u=410$~\kmsec~ the upstream flow speed, we get an Alfv\'{e}n Mach number of about 4, and the ratio of the downstream to upstream field is $B_d/B_u=3$. Both values point to a moderately strong  shock.


\textbf{Method (2):}

Figure~\ref{fig:global_shock}(b) shows an illustration of the four--spacecraft method. It assumes local planarity of the shock and a constant propagation direction and speed. When the spacecraft positions $R_i$ and the differences $\Delta t_{\alpha1}$ (with $\alpha=2,3,4$) of the shock arrival times are known, the shock normal $\mathbf{n_{s}}$ and speed $\mathbf{V_{s}}$ can be obtained from:
\begin{eqnarray}\label{equ:shock4}
\mathbf{V_s}=\nu \cdot \mathbf{n_s}, \\
   (\mathbf{R}_{\alpha}-\mathbf{R}_1) \cdot \mathbf{n_s}=\nu \cdot \Delta t_{\alpha i},
\end{eqnarray}
with $\nu$ being the norm of the shock normal speed. The results for this method are $\textbf{n}_s=(-0.907, -0.218, 0.361)$, or orientation angle $\theta_s=21^\circ$, $\phi_{s}=194^\circ$,  and $q=40^\circ$, $V_{s}=511$~\kmsecc  and $M_A=5.0$. We thus find a good agreement between this and the single-spacecraft method, with differences in the shock normal of only 8$^\circ$ and in the speed of 14~\kmsec. This means that the shock normal near Earth is well determined, indicating that the shock surface is reasonably close to planar on a scale  of $\approx$100~$R_E$.

\subsection{Implications for the Global Shock Shape and Structure}

We now discuss the individual shock parameters, summarized in Table~\ref{tab:shocks}, in view of the question: may we extrapolate features of the shock surface from a local to the global scale?

It was found in previous studies that even on a length scale of only a few $R_E$, shock normals can be inconsistent \cite[e.g.,][]{sza01}, pointing to non-planarity or corrugation of the shock surface on very small scales compared to the global extent. We can observe the shock \emph{S2} at a longitudinal spacecraft separation of 120\degreee, which corresponds to a length of $l_s \sim 2.1$~AU at the heliocentric distance of 1~AU (neglecting the non-spherical shape of the shock). The maximum separation of near-Earth spacecraft is of the order of a few hundred $R_E$, which is roughly a factor 100 less than  $l_s$. It thus becomes clear that for a corrugated shock surface on scales much less than $l_s$, it would be impossible to extrapolate the global shape of the shock from a single-spacecraft observation. However, \cite{sza01} also note that fast ejecta impose a more pronounced planarity to the shocks which they are driving as compared to slower ones. A prime example for such a ``well-behaving'' shock near Earth driven by a fast ejecta is discussed by \cite{fou07}. It may thus be expected that, for a fast compound stream like the structures following \emph{S2}, with speeds up to 800~\kmsecc at \STB and 600~\kmsecc at \emph{Wind}, it could indeed be possible to extrapolate the principal direction of the shock front from the single-point shock normals and see if it is consistent with the  ``real'' situation as determined by the multi-spacecraft observations. 

To this end, we may take a look again at Figure~\ref{fig:global_shock}(a). In particular, note the longitude of the shock normals, which flare away from the radial direction, towards the east at \emph{STEREO-B}, slightly west at \VEX and more pronounced to the west at \emph{Wind}. This is intuitively expected for a non-corrugated shock surface which is convex with respect to the Sun. 
Only \MES does not follow this pattern, which could be a hint that the shock surface at the outer flank is not driven strongly enough. An observer looking at near-Earth spacecraft would thus have correctly deduced, by looking at the longitude of the shock normal, that the apex of this shock is to the east of Earth. This means that the  principal direction of the shock can be correctly obtained in this case from single-point (using co-planarity) or multi-point (the four-spacecraft time-of-flight method) shock analysis of a shock driven by a fast ejecta in near-Earth space.


The configuration of the shock \emph{S2} seems to be altered by the presence of the previous ICME at the location of \emph{STEREO-B}, where the shock strength is lower, due to the higher field strength and lower density \citep{liu12}.  
 Otherwise, the strength of the shock is moderately strong for the full extent. The shock \emph{S2} is also everywhere oblique except at \emph{STEREO-B}, where it is quasi-perpendicular, which is consistent with expectations. In general, shocks are more likely to be quasi-perpendicular while propagating inside an ICME \citep{ric10perp}. At \emph{Wind} and \emph{VEX}, the shock has just reached the back of a previous  ejecta and seems not yet strongly altered in its configuration. From both methods we used in near-Earth space, the angle $q$ is still close to 45\degreee, as expected for a spherical shock propagating through a nominal Parker spiral field (cf. Figure~\ref{fig:global_shock}(a)). In contrast to the shock normal, the shock configuration is a function of the upstream field, which complicates the situation. Nevertheless, for this event it looks like the configuration of the shock as a function of  longitude behaves as it is expected from our general understanding of interplanetary shocks. 


In summary, the global shape of the shock \emph{S2} is clearly non-spherical, though convex with respect to the Sun as expected, and the longitude of the shock normals $\mathbf{n_s}$, follow a pattern consistent with this shape at three out four locations. The following relation is valid: if the longitude of $\mathbf{n_s}$ points east (west) of the radial direction, the shock apex is to the west (east) of the spacecraft. Only at the most western location at \emph{MESSENGER}, which represents its flank, the opposite is seen.  This also places the apex of the shock between \emph{STEREO-B} and \emph{VEX}. The configuration of the shock as a function of heliospheric longitude is complicated by the presence of previous ICMEs, but follows a pattern expected for observations at 1 AU.

\section{Flux Rope Analysis}

In this section we use different modeling techniques to extract more information on the observed large-scale magnetic flux ropes (MFRs), 
i.e., those parts of the ICME intervals which exhibit a smoothly rotating field over a large angle, in addition to higher-than-average magnetic field. These are the structures we have defined as \emph{M1/2/3} at the spacecraft \emph{STEREO-B}, \VEX and \emph{Wind}, and \emph{M1} at \emph{MESSENGER}. Our intention is to find matches between the model results to find out if the structures observed at a single spacecraft may be connected to each other and how the resulting global structures may look like. We use the same conventions for coordinate systems and orientation angles as for the shocks in the previous section.


\subsection{Flux Ropes Observed by \emph{Wind}}

 \begin{figure}[h]
\epsscale{1.2}
\plotone{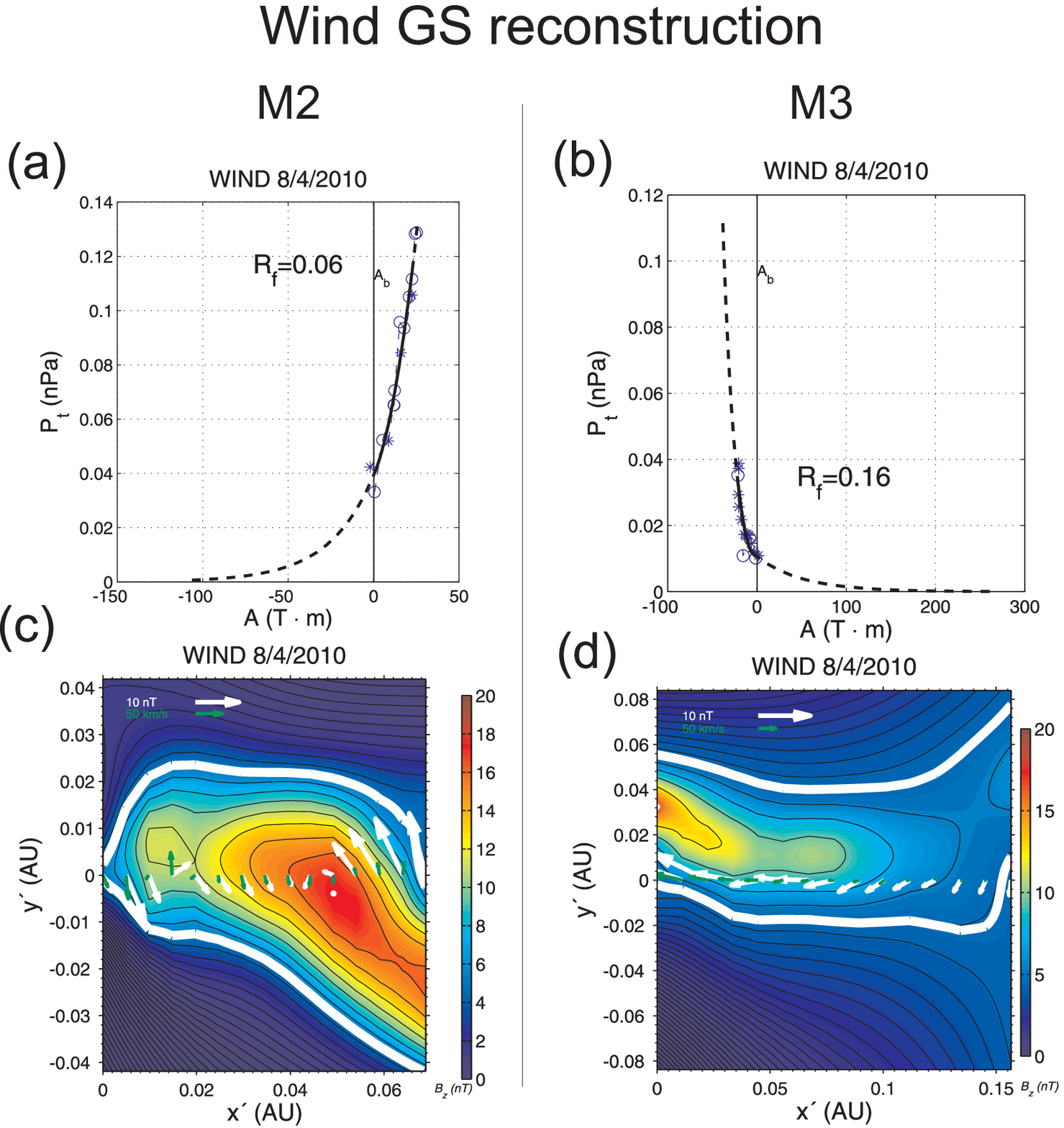}
 \caption{ (a, b) Transverse pressure relations and fitting functions (black solid and dashed lines). (c, d) Magnetic cloud cross-sections: the black contours are magnetic field lines in the $x'$-$y'$ plane, which is perpendicular to the MFR axis, which is pointing out of the paper plane. The spacecraft traverses from left to right, and the Sun is to the right \citep[for exact definitions see ][]{hu02}. The  axial component of the field, $B_z'$, is color coded. White arrows are magnetic field observations by \emph{Wind} in the $x'$-$y'$ plane, resampled to 15 points and used as initial values for the numerical integration. Green arrows are small residual flow velocities in the deHoffmann--Teller frame. The white contour corresponds to the value of the vector potential $A$ as indicated in panels (a, b) by a vertical line at $A_b$ and are defined as the boundary for the flux and current calculations (see Table~\ref{tab:flux}).}\label{fig:wind_gs}
 \end{figure}

An inspection of Figure~\ref{fig:wind}, showing the magnetic field components in GSE coordinates and proton bulk parameters, indicates that one MFR (labelled M1) may be identified before the arrival of shock \emph{S2}, and two distinct MFRs can identified at \emph{Wind} after the shock \emph{S2} and its following sheath structure, which is the region of ``shocked'' solar wind with high temperature, density and proton $\beta_p$. MFRs \emph{M2} and \emph{M3} are immersed in a declining flow speed ($V_p$) profile with two maxima which is typical for complex ejecta \citep{bur02}, and it also satisfies the definition of a compound stream. Thus, it is immediately clear that this flow is composed of several structures, and, as was shown in \cite{liu12}, these structures can still be treated as individual entities, also at the other spacecraft. Thus this stream is not classified as a ``complex ejecta'' in the sense of \cite{bur02}. \cite{liu12} suggested that structure \emph{M1}, whose back part seems to be ``caught'' by the shock just before arrival at 1~AU, could be the flank of a CME originating on the Sun on July 30, so all ICME signatures except the proton temperature are not well pronounced, and thus we will not discuss this structure further here.

The region \emph{M2} started on August 4 04:34 UT and it extended almost seamlessly until the front boundary of \emph{M3} arrived on August 4 10:07 UT, and \emph{M3} extended further to August 4 23:18 UT. Both MFRs were associated with a low variance magnetic field and a smooth rotation of the magnetic field direction, and the proton temperature in \emph{M3} is clearly low enough to classify this structure as a magnetic cloud \citep{bur81}. Also, $\beta_p$ showed a clear depression and the solar wind speed had a declining profile, the latter indicative of ongoing radial expansion when it passed \emph{Wind}. These are typical signatures of MCs at 1~AU, though its size of $\sim 0.15$~AU is not particularly large \citep[e.g.,][]{jia06}. 

In contrast, the structure \emph{M2} shows stronger maximum fields (19.4~nT compared to 14.1~nT for \emph{M3}), higher proton temperatures, an enhancement in density denoted by \emph{D1}, a smaller size ($\sim 0.1$~AU) and no radial expansion, which are all indicators of interaction with \emph{M3} after \cite{far04}. It could also be that the shock \emph{S2} passed through \emph{M2} before 1~AU, and led to additional compression and heating, which is consistent with what is seen in heliospheric images \citep{liu12}. At the leading edge of \emph{M3}, there is another density increase (\emph{D2}), which is stronger and of shorter duration. We do not think that it is the cause of  compression, but rather that it is possible filament material. We plan to discuss this observation in detail in a subsequent publication.

More specifically, the average speed of the expansion is defined as a half of the difference between the speeds at the leading and trailing edges of the ICME. For \emph{M3}, the measured speeds were 585~km~s$^{-1}$ and 504~km~s$^{-1}$, respectively, yielding an expansion speed of 41 km~s$^{-1}$. This is $\approx 30$~km~s$^{-1}$ less than typical for events observed at 1~AU \citep{jia06}.  On the contrary, \emph{M2} was associated with a positive speed gradient of 8 km~s$^{-1}$ from the trailing to leading boundary, thus it was even contracting slightly as it passed 1~AU, which could also explain its smaller size.

To estimate the local MC axis orientations and other parameters we used the method of Grad-Shafranov (GS) reconstruction \citep[e.g.,][]{hu02,liu08,moe09b,moe09, isa11}. This method assumes time-independence and invariance along one direction which is equated with the axis of the MC. The first assumption is approximately fulfilled for \emph{M2} as shown above and also for \emph{M3} since the radial expansion is in the order of 10$\%$ of the average speed for \emph{M3}. By numerical integration, a map of the MC magnetic field lines in a plane perpendicular to the axis can be created without assuming a particular shape for the cross-section. The technique could be successfully applied to both MFR intervals. The resultant plots are shown in Figure~\ref{fig:wind_gs} and the calculated values are again summarized in Table \ref{tab:flux}.

Figures~\ref{fig:wind_gs}(a) and (b) show the transverse pressure $P_t=B_z'^2/2\mu + p$ as a function of the vector potential $\textbf{A}$ for each flux rope. Note that the plasma pressure is calculated from the protons only. For the assumed symmetry, the vector potential may be expressed in the form $A(x, y) {\bf e_z}$, where $z$ is the axis direction and $\mathbf B$ is
then expressed as ${\bf B} = \nabla A \times {\bf e_z} + B_z {\bf e_z}$. The local MC axis orientation is found by constraining $P_t$ to be a single-valued function of $A$, i.e., having the same values along the in- and outgoing paths. To see how well this function can be fitted with a polynomial (of order 2, black solid line) with exponential tails (black dashed lines) which is used for the numerical integration, a fitting residue $R_f$ is defined \citep{hu04}. For \emph{M2}, the value of $R_f=0.06$ indicates a very good fit. For \emph{M3}, this function has to be quite strongly extrapolated, and the  fitting residue ($R_f=0.16$) is relatively high, which indicates that the method is not working well and that the magnetic field map and the results in Table~\ref{tab:flux} for this structure should be taken with care. This is also clear from looking at the magnetic field components, where it is not  straightforward to discern a clear full rotation of the field. Figures~\ref{fig:wind_gs}(c) and (d) show the magnetic field cross--sections as contour lines. The shape of a \emph{M2} appears to be rather deformed, possibly as a result of the interaction with \emph{M3}. The shape of the resulting cross--section of \emph{M3} is reminiscent of a glancing encounter; the strongest axial field is at the left border of the integration domain.

In summary, the \emph{Wind} observations of \emph{M3} are likely to reflect a glancing encounter with a large MFR \citep[e.g.,][]{mar07,moe10}. In contrast,   \emph{M2} shows a clear magnetic field rotation on roughly half the scale size (0.07 AU), which could be caused by a more central crossing of a smaller ICME and/or compression by \emph{M3}.

\begin{table*}[t]
\scriptsize
\begin{center}
\caption{Modeled Flux Rope Parameters at Each Spacecraft. }\label{tab:flux}
\begin{tabular}{lcccccccccccccc}
\tableline\tableline
FR  & Type & Start Time & $\Delta$t & $V$(\kmsec) & $B_{z} (nT)$ &$H$& $\theta_m$(\degreee) & $ \phi_m$(\degreee) & $D$ (AU) & $p$ (AU)& $\Phi_t$ ($10^{21}$~Mx)& $\Phi_p$ ($10^{21}$~Mx) & $I_z$ (MA)\\
(1) & (2) & (3) &(4) &(5) &(6) &(7) &(8) &(9) &(10) &(11) &(12) &(13) & (14) \\
\tableline
\emph{STB} M3 MV  & NES  &    Aug 3 21:50 & 9h:12m &    544  & 12.8 & R   &-54  & 43 &    0.120 & \ldots   & \ldots     & \ldots         &\ldots           & \ldots \\
\emph{VEX}  M3 MV  & -ES & Aug 2 19:51 &    11h:21m   & \ldots  &  17.7 & R & -24 & 75  &  0.14\tablenotemark{a}   & \ldots     & \ldots         &\ldots           & \ldots\\
\emph{Wind} M2 GS & SWN & Aug 4 02:50 & 5h:08m & 569 & 17.7  &  R & 23 & 251  & 0.069 & -0.004 &  0.071 & 0.383 &    228	\\
\emph{Wind} M3 GS & -WS & Aug 4 09:38 & 13h:40m &  544 & 16.6  &  L & -39 & 301 & 0.157 &  0.032 &  0.127 & 0.570 &   47 \\
\emph{MES} M1 MV & NWS & Aug 2 14:12 & 2h:29m & \ldots & 24.2  &  L & 13 & 328 & 0.035\tablenotemark{b} &  \ldots & \ldots & \ldots & \ldots \\
\tableline
\end{tabular}
\end{center}\tablecomments{
Column 1: spacecraft, flux rope (FR), type of method: GS=Grad--Shafranov \citep[after][]{hu02,moe09b}; MV = minimum variance  \citep[after][]{son67, bot98}. Column 2: type of field rotation determined by eye \citep[after][]{bot98,mul98}. Column 3: FR front boundary. Column 4: duration of the FR. Column 5: Norm of deHoffmann-Teller velocity. Column 6: for GS: maximum value of the axial field component, MV: maximum magnetic field along the spacecraft trajectory. Column 7: FR handedness. Column 8: Inclination of the FR axis to the ecliptic plane ($\theta$=[-90 ... +90]\degreee). Column 9: FR axis longitude ($\phi$=[0 ... 360]\degreee; sunward$= 0$\degreee; solar east$=90$\degreee). Column 10: GS: radial diameter $D$ of the FR perpendicular to axis, MV: diameter along the spacecraft trajectory. Column 11: impact parameter $p$, the closest distance to the FR axis (only GS). Column 12: the axial flux of the FR (only GS). Column 13: the poloidal flux for a FR length of 1~AU (only GS). Column 14: the axial current $I_z$ of the FR (only GS).
\tablenotetext{1}{Assuming a radial speed of 550 \kmsecc times the duration of the FR. For the visualization of this FR size in Figure~\ref{fig:global_flux}, an expansion to 0.223 AU was estimated according to a quasi-linear law by \cite{lei07} for the expansion of magnetic cloud diameters. The radial position of the FR in Figure~\ref{fig:global_flux} corresponds to the time of the shock arrival at \emph{Wind} (see the text).}}
\tablenotetext{2}{Assuming a radial speed of 600 \kmsec. For the size of this FR in Figure~\ref{fig:global_flux}, the \cite{lei07} expansion law yields 0.075 AU at the time of the \emph{Wind} shock arrival (see the text).}
 \end{table*}

\subsection{Flux Ropes at \emph{STEREO-B},   \emph{VEX} and \MES}

We now briefly discuss large-scale flux ropes observations at the other spacecraft. Modeling was not as successful as in near-Earth space, so we limit our discussion here. We also skip discussion of the flux rope \emph{M1} at \STB and \emph{VEX}, which follows shock \emph{S1} and originates from an eruption on July 30 \citep{liu12}. The main results are again summarized in Table 3.

By looking at the magnetic field components in Figures \ref{fig:stereob} and~\ref{fig:vex}, we find very similar rotations for the structure \emph{M3}, and at the relatively small separation of 18\degree between the two spacecraft it is quite clear that the same structure is detected at two different sites. At \STB it is a north-east-south (NES, right handed) flux rope with marginal rotation, classifying this as an MCL structure with an axial field pointing eastward and an axis close to the ecliptic plane. At \emph{STEREO-B}, the speed profile is also roughly declining, as would be expected for an expanding flux rope. At \emph{VEX}, one sees only the east-south (ES) part. Probably, this is a glancing encounter of an ICME flux rope which went to the north or south of the ecliptic. Note that non-force-free modeling gives consistent results in the orientation, yielding at both locations a low inclination to the ecliptic plane with eastward axial field \citep{web12}.

At first glance, \emph{M2} at \STB and \VEX can be seen as a distinct MFR with a well-pronounced field rotation. It can also be successfully modeled as a small left-handed flux rope with moderate inclination pointing southward of the ecliptic. However, at both spacecraft there is no sheath region, so the rope follows the shock immediately, and thus we think it is much more likely that this is just the shocked part of the \emph{M1} structure and not a separate ICME. The scenario of the shock \emph{S2} reaching the back of the CME on July 30 (connected to the structures \emph{S1/M1}) just before 1 AU is also consistent with heliospheric imaging \citep{liu12}.

At \emph{MESSENGER}, we could identify only one short interval where the field is relatively smooth and rotating. This follows the shock \emph{S2} and is an NWS rope, with the axis being close to the ecliptic plane and the axial field pointing in the $+T$ (RTN coordinates) or solar west direction.  

\subsection{Plotting the Global Flux Rope Structure}

In Figure~\ref{fig:global_flux}, the local orientations of those structures for which we could find an axial orientation as given in Table~\ref{tab:flux} are visualized by different viewing angles. Figure~\ref{fig:global_flux}(a) shows a view of the ecliptic plane, whereas Figure~\ref{fig:global_flux}(b) shows a viewpoint slightly tilted from solar west. Figure~\ref{fig:global_flux}(c) shows the viewpoint looking from the radial going through Earth toward the Sun. The orientations are visualized by cylinders with sizes derived from modeling as taken from Table~\ref{tab:flux}. However, in a manner similar to the correction we imposed for the shock surface,  we tried to reproduce the sizes and positions of the MFR at one ``snapshot'' reference point in time, for which we deliberately choose the \emph{S2} shock arrival time at the \WIN spacecraft ($t_{S2W}$). To this end, we moved the flux rope centers according to Equation~(1), but now using the average speeds of the flux ropes (see Table~\ref{tab:flux}, Column 5) to correct for the different arrival times at different heliocentric distances. As an example, the flux rope center arrives  at \emph{VEX} on August 3 01:30 UT, which is 15h35min ahead of the shock arrival time at \emph{Wind}. During this time, the \emph{M3} center moves 0.206 AU further away from the Sun  (assuming a uniform speed of 550 \kmsecc for \emph{M3}), and at the time $t_{S2W}$ it is positioned at $r_{M3V}$=0.933 AU, and this is the distance at which we place this MFR in Figure~\ref{fig:global_flux}. For \emph{STEREO-B}, the position of \emph{M3} in Figure~\ref{fig:global_flux} ($r_{M3V}=0.9381$ ~AU) is \emph{less} than $r_B$, because the center of the flux rope arrives later (August 4 02:26 UT) than the reference time $t_{S2W}$. In summary, at our reference time, the flux rope \emph{M3} had an almost identical heliocentric distance along the radials going from the Sun through the positions of \STB and \emph{VEX}.

We also expanded (or contracted) the MFR size from the time of  the observation of the MFR center at the spacecraft to the reference time, for all spacecraft except \emph{Wind}. Similar to the estimate for the maximum magnetic field strength of the ICME flows at 1 AU as function of heliospheric longitude (see Equation~(\ref{mylaw})  and Table~1), we employ a similar correction for the ongoing expansion of the MFRs. The average expansion of MCs in the inner heliosphere, which we again use as a rough proxy, was quoted by \cite{lei07} as
\begin{equation}\label{leitnerlawsize}
   D(r) =0.23 \; r^{1.14}, 
\end{equation}
with $D$ and $r$ in (AU). A similar approach as in Section 2.3 leads to 
\begin{equation}\label{mylawsize}
 D'=D_h (r'/r_{h})^{1.14}.
\end{equation}
This formula tells us the size to which an average MC would expand at any heliocentric distance $r'$ if the diameter $D_h$ of the MC is observed at a heliocentric distance $r_h$. As an example, for the  application to \emph{VEX} we use for $r'$ the estimated position of the flux rope center ($r_{M3V}$) from above and an estimated flux rope diameter at $r_V$ of $D_V=0.14$~AU (again assuming the MFR has a uniform speed of 550 \kmsec).  Equation~(\ref{mylawsize}) then leads to a diameter $D'= 0.186$~AU, which we use for the size of the pink cylinder at \VEX in Figure~\ref{fig:global_flux}. The result for \STB is a contraction from 0.120~AU to 0.106~AU, which again follows from the fact that the center of \emph{M3} hits \STB \emph{later} than $t_{S2W}$.

\subsection{Implications for Global Flux Rope Structure}

In summary, Figure~\ref{fig:global_flux} should be seen as the final result of our flux rope analysis. We note several interesting hints on the global flux rope configuration, following from this figure. We indicated the MFRs which seem to belong to the same large structure with similar colors: pink for \emph{M3} at \STB and \emph{VEX}, blue for \emph{M2} at \emph{Wind}, and red for \WIN \emph{M3} and \MES \emph{M1}. These associations follow quite naturally by looking at the chirality of the field and the axial field directions. In particular, \emph{M3} for \STB and \emph{VEX} cannot extend to Earth because of the different axial field direction, which is eastward for the eastern (pink) structure and westward for the western (red) structure. Also, the field chirality is the opposite, so we think that these associations are solid. Further, it could be that \emph{M1} at \MES belongs to the same structure as \emph{M3} at Earth, because the axial field directions and chirality (left handed) are similar. However, these observations are very separated,  in both space (0.6 AU radial distance, 49\degree heliospheric longitude) and time (about two days). While the separation in longitude is less than an often quoted limit of 60\degree for the flux rope extent in heliospheric longitude \citep{bot98}, \cite{web12} did not find connections between the two structures from SMEI density reconstructions. All this makes it difficult to tell if these observations sample the same flux rope or not. 

By looking at the longitude of the axes, best viewed in Figure~\ref{fig:global_flux}(a), it seems that imagining the flux rope as a bent tube on a large scale works relatively well for these events. However, Figure~\ref{fig:global_flux}(c) shows that the latitude of the axes for the pink and red tubes are quite inconsistent with each other -- cautioning that the red tube might be a different structure.  This might be a hint that an inclination of a flux rope should rather be viewed as a local parameter. It is, however, difficult to distinguish true axial distortions or a ``warped'' shape on a large scale from a possible failure of the methods we used for finding the axial orientation in the first place. One should be especially cautious of over-interpreting the orientation results if the field rotations inside the flux ropes are not well defined, as it is true here for the events at \STB and \emph{VEX}  (see also Al-Haddad et al., 2012, Solar Physics, in revision). However, \cite{web12} show fits of non-linear force-free models \citep{mul01} to these MFRs and they are quite consistent with the results presented here. Because different models give us quite similar answers here, we think that the picture presented in Figure~\ref{fig:global_flux} is a relatively robust one.

\subsection{Associating ICME Signatures to CMEs}

\begin{figure*}[h]
\epsscale{1.1}
\plotone{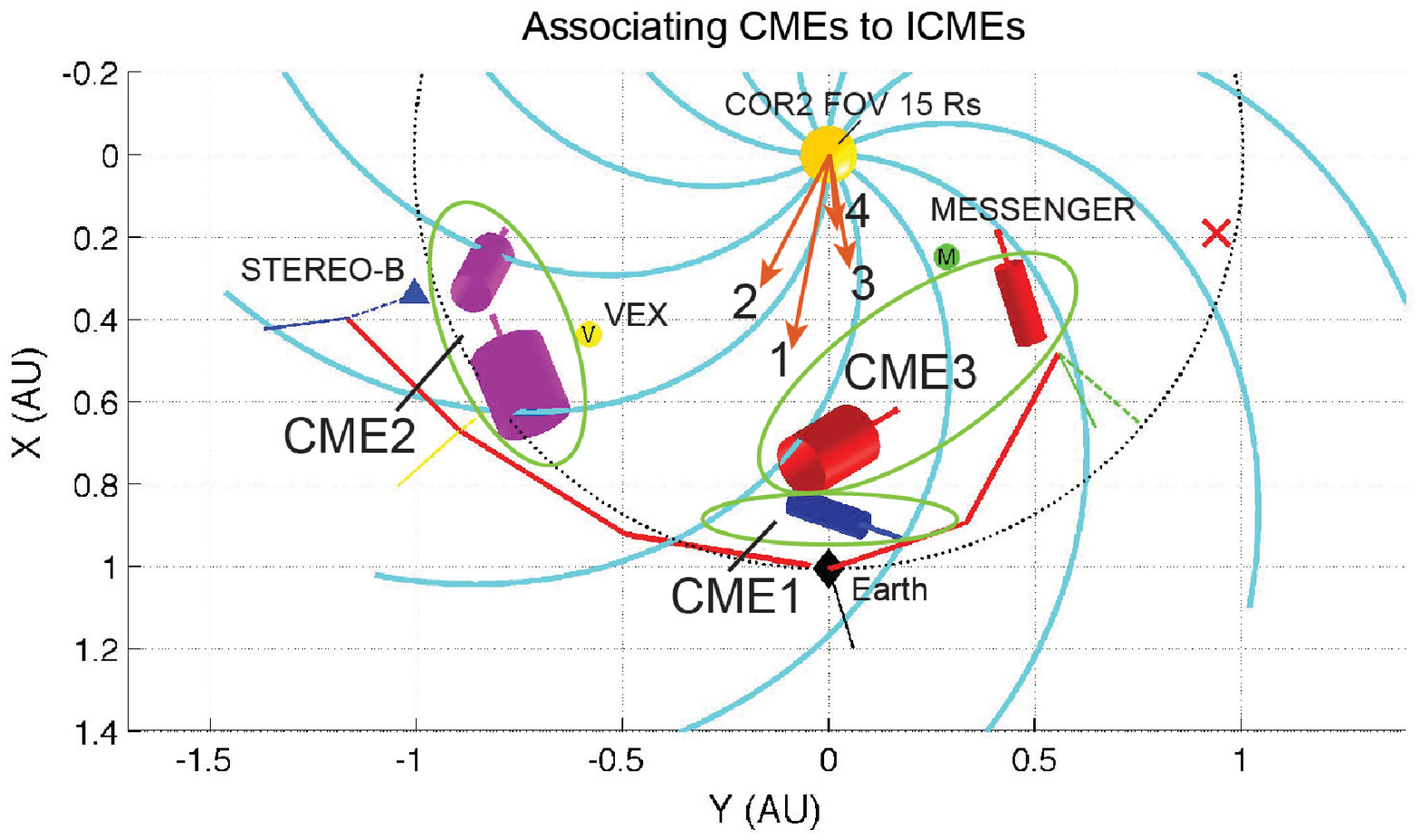}
\caption{View of the ecliptic plane visualizing flux ropes as cylindrical tubes and the shock surface \emph{S2} as a red solid line. The initial CME directions (orange arrows) are derived from \emph{STEREO}/COR2 by \cite{ods12}, with a shorter arrow indicating a later eruption. The \emph{STEREO}/COR2 maximum field of view (FOV) of 15 $R_{\odot}$ is shown by a yellow sphere. Green ellipses suggest which flux rope observations likely originate from the same CME. }\label{fig:associate}
\end{figure*}

In Figure~\ref{fig:associate} we try to relate the observed multi-point flux rope structures to the measured CMEs in coronagraphs. Similar to Figure~\ref{fig:all}(a), orange arrows indicate initial directions for the four CMEs on 2010 August 1 as given by an elliptical forward modeling technique applied to both \emph{STEREO}/COR2 coronagraphs \citep{ods12}. We think these results represent well the conclusions of other papers to determine the directions of these CMEs. For example, \cite{tem12} found a range for the direction for CME2 from E6 to E41, and \cite{ods12} find E25 for CME2. \cite{har12}, discussing the propagation phase in \emph{STEREO}/HI for CME2 (in their paper called CME L), find a direction between E0 and E40, which is also consistent with E25. By using green ellipses, we group flux rope observations which likely can be traced back to a similar CME.

\begin{figure*}[h]
\epsscale{1.1}
\plotone{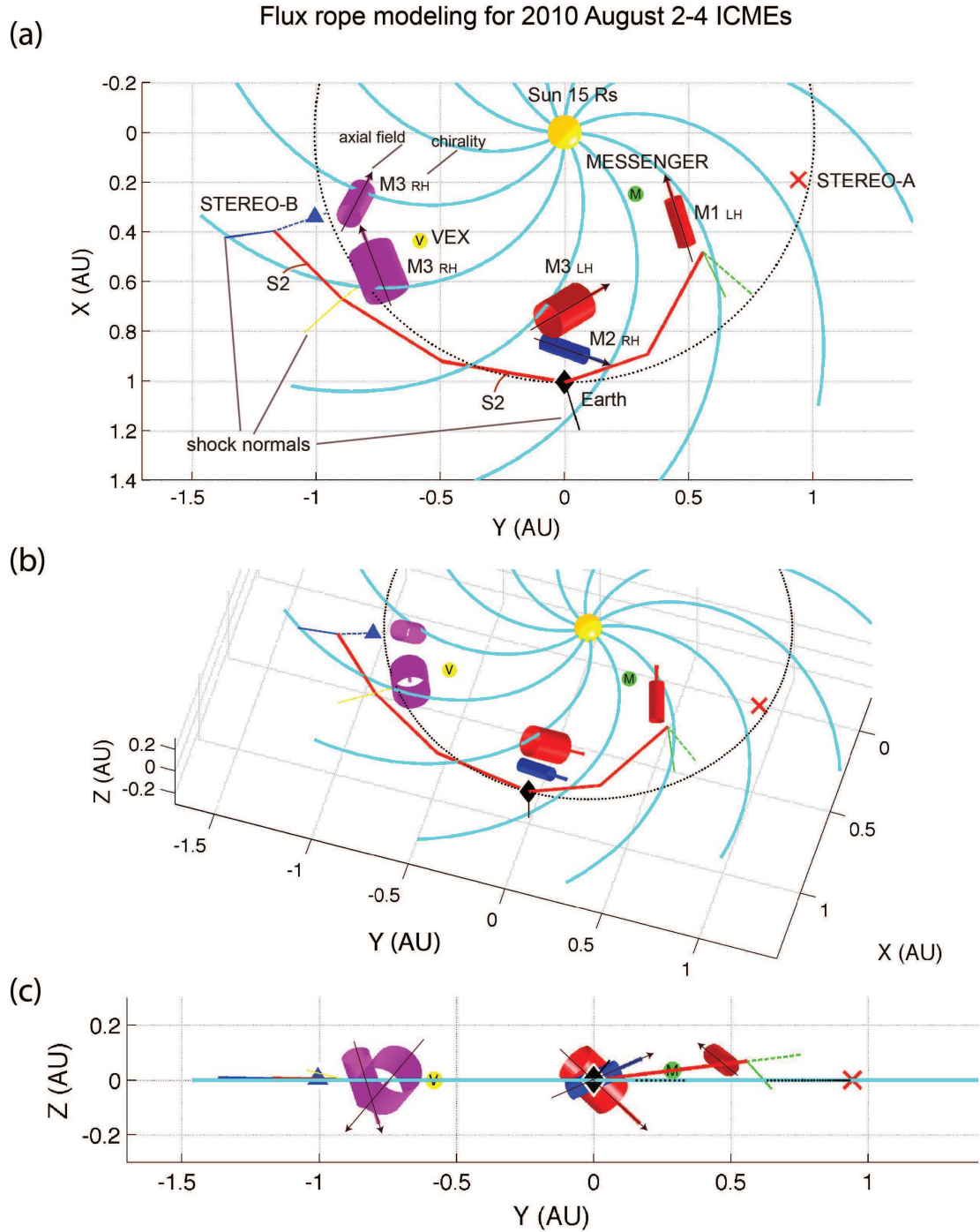}
\caption{Global configuration of magnetic flux ropes. Cylinders indicate sizes and orientations of large-scale magnetic flux ropes which could be successfully modeled. The red solid line is the approximate shape of the shock front \emph{S2}. The positions and sizes of the MFRs are corrected to approximate the global configuration of the ICMEs at a snapshot in time, which we choose as the shock arrival time at the \WIN spacecraft (2010 August 3 17:05). (a) View of the ecliptic plane from north. (b) Side view tilted to solar west. (c) Looking along the radial from the Sun to Earth. (An animation of this figure is available in the online edition of the \emph{Astrophysical Journal}.) }\label{fig:global_flux}
\end{figure*}

First, it is quite straightforward to associate the \emph{M3} structure at \STB and \emph{VEX} (pink tubes in Figures~\ref{fig:global_flux} and~\ref{fig:associate}) with the fast CME2, mainly because CME2 was the one with the most eastward initial direction. It also had a significant northward direction of $\approx$~N20 \citep{tem12,ods12}, which should have contributed to a glancing encounter for these spacecraft, resulting in the rather unpronounced magnetic field rotation at \STB and \emph{VEX}. Note that \cite{web12} do not find a connection between these two structures from SMEI density reconstructions, but it follows from the magnetic field modeling that they should belong to the same large-scale structure.

It is intriguing that \emph{M3} was seen at \STB and \VEX located 46\degree and 28\degree to the east of the most likely initial direction for the CME2, respectively. Given the 60\degree limit for widths of MCs \citep{bot98} noted earlier, one would expect \VEX to encounter the edge of CME2 and that it misses \STB completely. This flux rope is not observed at Earth, though it is 25\degree west of the CME2 initial direction, which is closer than \emph{VEX}. Explanations for this could be that CME2 propagated more easily eastward into the depleted density region by the 2010 July 30 CME, or that the interaction with CME1 around 30--50~$R_{\odot}$ \citep{liu12,tem12} may have led to a deflection to the east for CME2 and to the west for CME1 (see also Lugaz et al. 2012, ApJ, in press, reaching similar conclusions for a different event). 

Indeed, at Earth, the \emph{M2} flux rope (blue), which does not have a counterpart elsewhere, might correspond to CME1, which had an initial direction also slightly to the east, but closer to Earth (see again Figure~\ref{fig:associate}). Furthermore, if the strong density peak \emph{D2} at the front boundary of \emph{M3} is indeed filament material, then \emph{M3} (the red tube) would correspond to CME3, which was almost co-temporal with CME2 \citep{sch11, har12} and associated with a filament eruption around disk center in the northern hemisphere \citep{sch11,li11b, har12}. 
 As discussed in the previous section, while it makes sense to associate the only clean flux rope observed at \emph{MESSENGER} with \emph{M3} at \emph{Wind}, the observations are too far separated in space and time to make a definitive connection.  The weak CME4 did not seem to leave any strong trace in the in situ data, or, alternatively, could be responsible for the \emph{MESSENGER} \emph{M1} flux rope.

In summary, for three out of four CMEs we could find reasonable connections to MFRs observed in situ. For at least one CME, a deflection during heliospheric propagation seems to be needed to explain the connection between the observations in coronagraphs and in situ. We have avoided associating the shocks with specific CMEs, because, for an extended shock surface like \emph{S2} spanning $> 120$\degree in heliolongitude, different CMEs may drive it at different frontal locations.

\section{A two-step geomagnetic storm}

In this last section of the paper, we will take a closer look on this compound stream as observed near Earth, to find out which parts of it were responsible for producing geomagnetic activity. 

 \begin{figure*}[h]
\epsscale{1}
\plotone{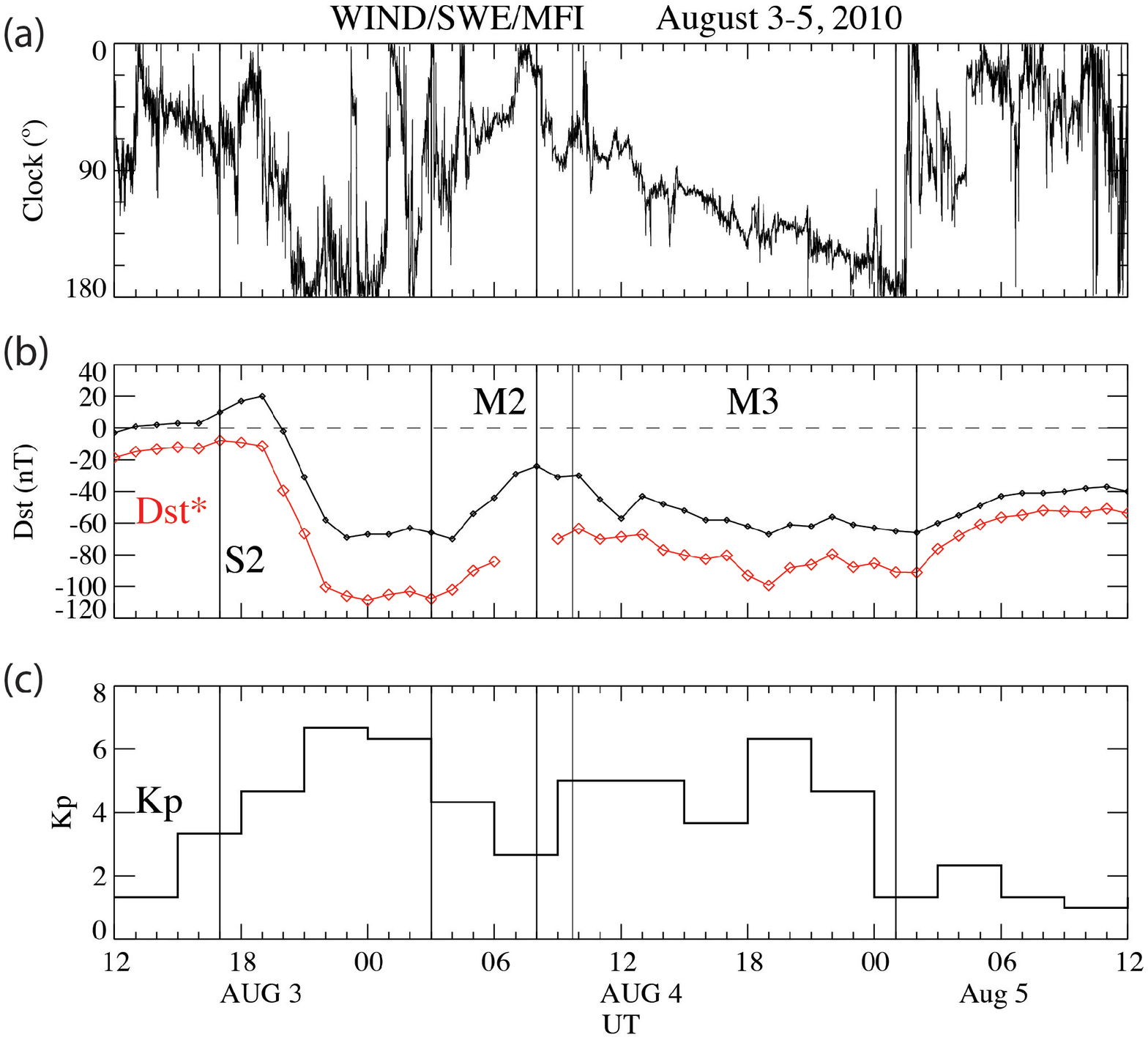}
 \caption{(a) Clock angle in the GSM $YZ$ plane of the interplanetary magnetic structure (0 ecliptic north, 180 ecliptic south), the (b) real time \emph{Dst} index at 1 hr temporal resolution (black) and the corrected \emph{Dst}* index (red), and the \emph{Kp} index at 3 hr resolution. Vertical guidelines label shocks and magnetic structures.}\label{fig:geoeffects}
 \end{figure*}

Figure \ref{fig:geoeffects} plots from top to bottom (a) the clock angle of the interplanetary magnetic field (IMF), i.e., the polar  angle in the GSM $YZ$ plane, (b) the \emph{Dst} index at 1 hr temporal resolution, and (c) the \emph{Kp} index at 3 hr resolution, obtained from the  World Data Center for Geomagnetism in  Kyoto. The polar angle is defined as 0\degree for a purely northward pointing IMF, 90\degree for an eastward oriented, and 180\degree for a purely southward pointing field.
 The 1-hr \emph{Dst} index is calculated from four low-latitude magnetogram stations and it monitors the strength of the symmetric equatorial ring current. In Figure \ref{fig:geoeffects}(b) we have also plotted the pressure-corrected \emph{Dst} index (\emph{Dst}*) where the effect of the magnetopause Chapman-Ferraro currents is removed according to \cite{bur75}.  This correction assumes a Chapman-Ferraro scaling and is not appropriate when the response of the magnetosphere to external forcing -- i.e., to the interplanetary motional electric field (IEF) -- becomes non-linear and approaches saturation \citep{obr00, sis02}. Saturation typically sets in at values of the IEF $\sim$ 6--8 mV m$^{-1}$ \citep{hai03, mue05}. In our case, IEF is less than 6~mV~m$^{-1}$ most of the time (not shown) and thus a linear response may be assumed so that  this correction should be valid. The \emph{Kp} index is calculated from 13 subauroral stations and is thus more sensitive to high-latitude activity than \emph{Dst}.

Figure~\ref{fig:geoeffects}(b) shows that there are two main decreases in \emph{Dst}*, with one significant recovery in between. The first depression was the strongest. The arrival of the shock and the following brief region of northward IMF and increased dynamic pressure were seen as a positive excursion in \emph{Dst}*, which is known as a sudden impulse (SI). After August 3 19 UT, the IMF turned southward and \emph{Dst}* started a rapid decrease. IMF was primarily negative throughout the sheath region and  \emph{Dst}* remained below -100 nT until August 4 03 UT after which it started to recover. The leading edge of \emph{M2} arrived at the magnetopause around August 4 04:30 UT, and thus the first \emph{Dst}* depression was evidently caused by the turbulent sheath region, which, however, lead to a relatively steady southward pointing field, which could be a consequence of the IMF draping around the ejecta in the sheath \citep[e.g.,][]{gos87,liu08c}.  The \emph{Dst}* minimum during this first depression was just below -100~nT at August 4 00 UT, thus it is classified as an intense magnetic storm after \cite{gon94} The \emph{Kp} maximum of 7$^{-}$ was reached on August 3 21--24 UT. During the second peak the \emph{Kp} maximum was 6$^+$ from August 4 18--21 UT.

During the passage of \emph{M2}, the GSM $B_z$ component was mainly  northward and \emph{Dst}* recovered.  A few hours after the arrival of \emph{M3} at the magnetopause IMF turned southward and \emph{Dst}* started its second decrease. Now, the IMF maintained steady southward orientation for more than 12 hr, keeping \emph{Dst}* low. However, the magnitude of the southward component was only about 5 nT and no severe activity followed. \emph{Kp} followed the same pattern as \emph{Dst}*.

As described above, the analyzed compound stream caused \emph{Dst}* to develop in two distinct steps. Such two-step magnetic storms \citep{kam98} are related to the double-structured southward IMF, caused typically by a shock-driving south-north (SN) cloud, with the shock enhancing previously negative $B_z$ upstream fields.  In such a case, the sheath fields initiate the storm that is further intensified by the following southward fields in the leading part of the cloud \citep{vie01}. Shock-driving north-south (NS)-clouds may also lead to a two-step storm, but with a longer separation between two consecutive \emph{Dst}* excursions and stronger recovery in between. If an NS-cloud has sufficiently long duration, the ring current may decay during the first part of the cloud close to the quiet-time levels, and two separate magnetic storms follow (according to \cite{kam98} this happens when the ratio of the magnitudes of the \emph{Dst}* recovery and the first \emph{Dst}* depression is larger than 0.9). However, the two-step storm analyzed in this paper was caused by the complex combination of a sheath region and two interacting MFRs. 

As the majority of flux ropes are expected to have NS field rotation during the rising phase of the current solar cycle 24 \citep[e.g.,][]{bot98, mul98, li11}, it is interesting to ask how strong the ring current response would have been if \emph{M2} had NS-polarity instead of SN. To estimate this we used the modified Burton's $Dst$* prediction scheme \citep{bur75, fen98, obr00b} for the interval of the compound stream with the sign of $B_z$ reversed during \emph{M2} (data not shown). In this reversed -$B_z$ case, the compression by \emph{M3} coincided with the southward fields in the end part of \emph{M2} and the following storm had three distinct $Dst$* excursions, with the strongest depression occurring during the passage of \emph{M2}. However, the $Dst$* minimum of the storm was only slightly stronger than when \emph{M2} had original polarity (-94 and -70 nT for the reversed and real -$B_z$ \emph{M2} cases, respectively; note that the $Dst$* model underestimates the strength of the storms) and the magnitude of the third depression remained about the same (-64 nT and -58 nT, respectively). Although compound streams often cause intense $Dst$* storms (see Introduction) it is not clear how much pre-existing ring current affects the storm strength \citep[e.g.,][]{koz02}.


The studied event is also a good reminder of the importance of the sheath fields in generating magnetic storms \citep{tsu88,hut02}. The strongest \emph{Dst}* depression was caused by the sheath region, and not the magnetic flux ropes. The magnetic field magnitude in \emph{M2} was high as a result of interaction with \emph{M3}, but the $B_z$ was mainly northward causing the recovery of the \emph{Dst}*. Concerning \emph{M3}, we already discussed that it is likely that the Earth made only a glancing encounter through this structure. As a consequence, the magnetic field magnitude was overall low and the following geomagnetic activity was not strong.

\section{Conclusion}

The series of events from 2010 July 30 to August 1 gives us a rare opportunity to ``see beyond'' single-spacecraft, in situ observations of ICMEs. We are actually unaware of any other series of events which was covered this well by four in situ spacecraft at almost equal longitudinal separations spanning 120\degree in longitude, which is a third of the heliosphere, though of course constrained to the ecliptic plane. As could be expected for a series of four coronal mass ejections heading clearly towards four out of five spacecraft from which we had the chance to obtain the in situ data, we found a plethora of physical phenomena such as shocks, flux ropes and interactions between various ejecta at these four locations, and attempted to link them into a global picture.

We summarize our main findings as follows.
\begin{enumerate}
\item A shock propagating partly into a previous ICME can lead to alterations of the shock structure; a similar conclusion was reached by \cite{liu12}. Additionally to this study, we have derived the approximate shape of the shock, showing it to be clearly non-spherical with respect to the Sun, though convex as expected. This non-spherical shape and a weakening of the shock at the eastern flank is caused (1) by parts of the shock propagating into a medium carved out by a previous ICME, and (2) a higher initial CME speed of the eruptions on 2010 August 1 on the eastern solar hemisphere. From the shock normal orientations longitudes, the global shape can be reasonably extrapolated from single-spacecraft observations. This means that the longitude of the shock normal is pointing east (west) of the radial direction to the Sun for spacecraft positioned on the eastern (western) flank of the shock, in three out of four shock normals (this was not true for the shock at \emph{MESSENGER}).   

\item Concerning modeling of the magnetic flux ropes (MFRs), we found a relatively good consistency of the results from applying several methods \citep[including those of ][]{web12} to the MFRs at all in situ locations with the classical picture of ICMEs as large-scale, bent magnetic tubes (Fig. \ref{fig:global_flux}(a)). In particular, we were able to associate \emph{M3} at \STB and \VEX as being parts of the same structure by looking at the chirality of the field and the axial field directions, at a longitudinal separation of $\approx 18$\degreee; in this way, it is also possible to associate \emph{M3} at Earth with \emph{M1} at \emph{MESSENGER}, though these observations are too separated from each other to make a definitive connection. While the longitudinal picture is consistent with the classical paradigm, looking at the inconsistent inclination of the flux ropes to the ecliptic plane (Figure \ref{fig:global_flux}(c)) might point to alterations of the flux rope picture as ``warped'' rather than straight tubes. One may imagine holding a deformable ``pool noodle''\footnote[1]{For this term we would like to give credit to Volker Bothmer who first introduced it at a STEREO meeting in Dublin in 2010.} in one's hands -- if deformed from its original circular shape, the inclination to a given plane varies along the axis.  We propose that we found some evidence for this axial ``warping'' in these events and thus, an inclination of a flux rope to the ecliptic plane may be viewed rather as a local quantity than a global one \citep[see also][]{far11}.

\item A shock propagating into a previous eruption can give rise to a rotating magnetic field which can be perfectly modeled with various methods as a small cylindrical flux rope, but could be mistaken for a distinct flux rope originating from a CME (\emph{M2} at \STB and \emph{VEX}). 

\item This series of events does not qualify as a complex ejecta as defined by \cite{bur02} because individual flux rope structures can still be clearly identified and modeled at all spacecraft. However, some of the parameters have been significantly altered in the interactions that did occur, in accordance with \cite{far04}. At \emph{Wind}, in particular, a shock propagated through the preceding MFR, and/or ongoing energy and momentum transfer between the MFRs lead to heating and compression of the plasma \cite[see also][]{liu12}.

\end{enumerate}

In conclusion, we think that this series of events, as complex as they are at first glance, may qualify very well as a benchmark test case for space weather prediction models. In numerical simulations for forecasting CMEs, the outputs of the model runs should be consistent with the arrival times of the shocks and flux ropes, the number of MFRs at each spacecraft and the general profiles of plasma and magnetic field. First attempts in this direction using cone-shaped ejections were extremely encouraging \citep[]{wu11,ods12}, but need to include the flux rope structures in future simulations. This would make it possible to attempt predictions of the magnetic field components inside the ICME, and concerns in particular the ever elusive $B_z$  component (in GSM coordinates) which can be directly linked to the strength of the geomagnetic storm \citep[e.g.,][]{tsu88}. Situations like the one described in this paper may be expected to be more common during solar maximum. Sympathetic eruptions naturally lead to compound streams composed of several CMEs, altering ICME properties in distinct ways, sometimes producing more geo-effective flows \citep[e.g.,][]{zha07}.

New possibilities for serendipitous observations such as these are expected for 2015, when the \ST  spacecraft will ``cross'' on the other side of the Sun as seen from Earth. With two additional spacecraft operating in the inner heliosphere at the moment,  \MES at Mercury and \VEX at Venus, and the Sun then in its declining phase of the cycle, it might be possible to ``catch'' more ICME events on a global scale such as the one presented in this paper. However, these future \ST  directed events might lack the three-dimensional imaging capabilites which can be used for the events in early 2010 August, which is another reason why we think these events are a truly outstanding test case for the space weather community.

\acknowledgments

We thank all the people involved in building and maintaining the great variety of instruments from which data was used to make this work possible. This work would also not have been possible without the enthusiasm of the attendees of the three workshops held in early 2011. We additionally thank Paulett Liewer, Karel Schrijver, and Volker Bothmer for enlightening discussions. This research was supported by a Marie Curie International Outgoing Fellowship within the 7th European Community Framework Programme. C.M., M.T., and T.R. were supported by the Austrian Science Fund (FWF): P20145-N16, V195-N16. The presented work has received funding from the European Union Seventh Framework Programme (FP7/2007-2013) under grant agreement no. 263252 [COMESEP]. It is also supported by NASA grants  NAS5-00132, NNG06GD41G, NNX08AD11G, NNX10AQ29G,  NNX08AQ16G, and NSF grant AGS-1140211. Work at the University of California, Berkeley, was supported from \emph{STEREO} grant NAS5-03131. D.F.W. was supported by Navy contracts N00173-07-1-G016 and N00173-10-1-G001. We also acknowledge the use of \WIN data provided by the magnetometer and the solar wind experiment teams at GSFC, and thank the centers for geomagnetism in Kyoto and Potsdam for providing the \emph{Kp} and \emph{Dst} indices.

\newpage

\newpage





\clearpage




\clearpage

\clearpage






\end{document}